\documentclass[aps,prb,showpacs,preprintnumbers,twocolumn,floats]{revtex4-1}%
\usepackage{graphicx}
\usepackage{amsmath}
\usepackage{epstopdf}
\usepackage{amsbsy}
\usepackage{color}
\usepackage{dcolumn}
\usepackage{bm}
\usepackage{hyperref}
\usepackage{amsfonts}
\usepackage{amssymb}%
\providecommand{\U}[1]{\protect\rule{.1in}{.1in}}

\graphicspath{{./figures/}}

\begin{document}
\title{Robust determination of superconducting gap sign changes via quasiparticle
interference}
\author{P. J. Hirschfeld$^{1}$, D. Altenfeld$^{2}$, I. Eremin$^{2,3}$, and I.I.
Mazin$^{4}$ }
\affiliation{$^{1}$Department of Physics, University of Florida, Gainesville, Florida
32611, USA}
\affiliation{$^{2}$ Institut f\"ur
Theoretische Physik III, Ruhr-Universit\"at Bochum, D-44801 Bochum, Germany}
\affiliation{$^{3}$ Kazan Federal University, Kazan 420008, Russian Federation}
\affiliation{$^{4}$Code 6393, Naval Research Laboratory, Washington, DC 20375, USA }
\date{\today}

\begin{abstract}
Phase-sensitive measurements of the superconducting gap in Fe-based
superconductors have proven more difficult than originally anticipated. While
quasiparticle interference (QPI) measurements based on scanning tunneling
spectroscopy are often proposed as defnitive tests of gap structure, the
analysis typically relies on details of the model employed. Here we point out
that the temperature dependence of momentum-integrated QPI data can be used to
identify gap sign changes in a qualitative way, and present an illustration
for $s_{\pm}$ and $s_{++}$ states in a system with typical Fe-pnictide Fermi surface.

\end{abstract}

\maketitle


There is considerable indirect evidence, but currently no ``smoking gun",
indicating that most, if not all, of the Fe-based superconductors condense
into a superconducting state that has $A_{1g}$ symmetry but changes sign
between electron and hole pockets. This is indeed the prediction of spin
fluctuation theory\cite{HKM_ROPP11} that follows from the simple argument that
the pairing interaction, proportional to the magnetic susceptibiltiy, is
repulsive and peaked at a wave vector $\mathbf{Q}$ which nearly nests these
pockets. The simplest version of this state, conventionally referred to as the
$s_{\pm}$~state, is isotropic on each Fermi surface pocket with opposite signs
for electron and hole sheets\cite{Mazin08}. The actual gap function realized
in Fe-based SC is in many cases thought to be highly anisotropic and even
possess nodes, but may still be considered $s_{\pm}$ as long as the average
sign on electron pockets is opposite that on hole pockets.

Arguments against the $s_{\pm}$ picture have been given as well. If orbital
fluctuations dominate spin fluctuations, a state with equal sign on all
pockets, denoted $s_{++}$, is favored\cite{Kontani10}. In addition, when either hole or
electron pockets disappear from the vicinity of the Fermi level, $d$-wave
pairing may be enhanced\cite{Maiti2011}. Thus, as in the cuprates, the
determination of the gap symmetry and structure is important as a clue to the
underlying mechanism of superconductivity. As discussed in Ref.
\onlinecite{HKM_ROPP11}, this task is not as straightforward as in the
cuprates due to the multiplicity of Fermi surface sheets.
While many tests of $s_{\pm}$ pairing symmetry have been proposed, all seem to
rely on specialized theoretical assumptions or are applicable only to certain
systems. The phase-sensitive probes that proved so decisive in the
identification of $d$-wave symmetry in the cuprates are much less useful in
the Fe-based superconductors, due to the fact that both $s_{\pm}$ and $s_{++}$-wave symmetries belong to the same $A_{1g} $ irreducible representation,
and the difficulty of fabricating junctions with controlled properties.
A general \textit{qualitative} test of another type that could distinguish an
$s_{\pm}$ pair state from others would be extremely useful.

One promising technique in this regard is quasiparticle interference, or
Fourier transform scanning tunneling microscopy (FT-STM). This probe measures the wavelengths of
Friedel oscillations caused by disorder present in a metallic or
superconducting system, which in principle contain information on the
electronic structure of the pure system. These wavelengths appear in the form
of peaks at particular wavevectors $\mathbf{q}(\omega)$, which disperse with
STM bias $V=\omega/e$. There is no reliable quantitative theory of
quasiparticle interference, in general because the sources of disorder, exact impurity potentials,  and  ${\mathbf{k}}%
$-dependence of the tunneling matrix elements are unknown. However, the positions of the
$\mathbf{q}$ do not depend on these effects and are related only to the
electronic structure, including the superconducting gap function.

The notion that subsets of these $\mathbf{q}$ that connect gaps of equal or
opposite sign on the Fermi surface can be enhanced or not according to the
type of disorder was introduced by Nunner et al.\cite{Nunner2006}.
Pereg-Barnea and Franz\cite{PeregBarnea2008} then proposed that a disordered
vortex lattice should behave like a set of localized order parameter
suppression scattering centers, and could be used to introduce disorder in a
controlled fashion with an external field. This experiment was performed on
the cuprate superconducting compound, Ca$_{2-x}$Na$_x$CuO$_2$Cl$_2$, by Hanaguri {\it et al.}\cite{Hanaguri2009}, who showed that
certain quasiparticle interference (QPI) $\mathbf{q}$ were indeed enhanced or suppressed by the field, in a
manner  apparently consistent with $d$-wave pairing.

This experiment was attempted in the Fe-based superconductors on a a Fe(Se,Te)
superconducting sample near optimal doping by Hanaguri \textit{et al}
\cite{Hanaguri2010}, who identified three interband scattering wave vectors $\mathbf{q}$: one  associated with
hole-electron scattering (the smallest momentum), and two associated with two
different electron-electron scatterings. The last two features were enhanced
in the presence of magnetic field with respect to the first one, which led the authors to conclude that
the hole and electron pockets have opposite signs of the order parameter, as
expected in the $s_{\pm}$ model\cite{Wang2010,Akbari2010}. Some caveats
regarding the interpretation, associated with the fact that two of the wave
vectors coincide with Bragg peaks, were discussed in Ref.
\onlinecite{HKM_ROPP11},  and there are other issues that we consider
below. More recently, Chi et al.\cite{Chi_etal_2014} measured QPI on LiFeAs in
zero external field, studying the bias dependence of a large $\mathbf{q}$ peak
corresponding to interband and a small $\mathbf{q}$ peak corresponding to
intraband scattering, arguing that upon moving from large to small bias, the
fact that one was suppressed while the other enhanced could be consistent only
with $s_{\pm}$ pairing. This is also an argument of the qualitative type, but
because the actual intensity dependence of the peaks as a function of bias was
in fact nonmonotonic, and because there is currently no well-founded theory
to support this conclusion, the interpretation is uncertain.

In fact, while QPI measured by FT-STM is in principle a
very powerful qualitative tool, it is nearly useless as a quantitative one.
This is because, while the poles of the response function ($\mathbf{q}$-spot
positions) are universal, the weights are not, but depend on the type,
strength and range of the scatterers (not generally known), as well as the
details of the electronic structure (including the gap), and the $z$-position at which the local density of states (LDOS) is
calculated. The actual pattern calculated in even the most sophisticated
theories then generally bears little detailed resemblance to the measured STM
Fourier intensities (for a discussion of some of the causes see
Ref.\onlinecite{Kreisel2015}), and identifying the symmetry or structure of
the gap function by a comparison of the intensity patterns is similar to
determining the age of a model from a Picasso cubist painting.

The situation is further exacerbated by advanced mythology that has formed
over years about QPI. A part of this mythology involves belief that the QPI
spectra are defined by the geometry of the Fermi surface in a nesting-like
way, another part that in the superconducting state these spectra are
proportional to coherence factors, and that magnetic and nonmagnetic
impurities have opposite effect on QPI. While this mythology has been
partially rectified in some papers, it has not become generally appreciated,
and some myths still persist.

A systematic summary of the established facts is badly needed, and we shall
provide it later in the paper. But, the main purpose of this paper is to find
a clean qualitative way to design a QPI experiment capable of distinguishing
$s_{\pm}$ from other pairings, and in general to identify sign-changing gaps
in an unconventional superconducting system. We argue that the best way is to
measure the intensity of integrated interband scattering peaks in the Fourier
transformed density of states, antisymmetrized with respect to STM bias, as a
function of temperature.
Then the integrated weights of the set of $\mathbf{q}$ corresponding to
sign-changing scatterings display a strong enhancements only for this channel; furthermore these
qualitative distinctions are robust against the strength of the scattering. We
conclude that while the general theory of QPI in unconventional
superconductors has many pitfalls, this particular consequence of pairing sign
change is robust and can be used to unambiguously identify $s_{\pm}$ pairing
if this type of distinction between intra- and inter-band pairing $\mathbf{q}%
$-peaks is observed.

\section{Formalism}

Here we present a model that captures all the qualitative features of the
general 2-band case, but allows analytical evaluation of the density of
states. We are interested in the density of states in a 2-band superconductor
of isotropic $s_{++}$ or $s_{\pm}$ type in the presence of disorder of various
kinds. Here, following the tradition, we use the the words \textquotedblleft%
1-band\textquotedblright\ and \textquotedblleft2-band\textquotedblright\ for
materials with arbitrary complex normal electronic structure, but with the
superconducting order parameter $\Delta$ that can be approximated by a single
value for all states on the Fermi level, or by two different values, depending
on the location in the Brillouin zone.

\subsection{1-band problem}

We first remind the reader of the 1-band problem. We assume a random
distribution of $N_{I}$ point-like impurities at sites $\mathbf{R}_{i}$ with
$i=1\ldots N_{I}$. The LDOS can be formally
decomposed $\rho(\mathbf{r},\omega)=\rho_{0}(\omega)+\delta\rho(\mathbf{r}%
,\omega)$ where $\rho_{0}$ is the DOS of the homogeneous superconductor, and
$\delta\rho$ is the local change in the DOS due to disorder given exactly by
\[
\delta\rho(\mathbf{r},\omega)=-\frac{1}{\pi}\operatorname{Im}\sum
_{i,j=1}^{N_{I}}\left[  \hat{G}^{0}(\mathbf{r}-\mathbf{R}_{i})\hat{T}_{ij}%
\hat{G}^{0}(\mathbf{R}_{j}-\mathbf{r})\right]  _{11}%
\]
where the $\omega$-dependence is suppressed for simplicity, $\hat{T}(\omega)$
is the $2N_{I}\times2N_{I}$ many-impurity $T$-matrix (the factor of two is due
to spin), $\hat{G}^{0}(\mathbf{r},\omega)$ is the bare advanced electron
Green's function $\hat{G}^{0}(\mathbf{r},\omega)=\sum_{\mathbf{k}}\hat{G}%
^{0}(\mathbf{k},\omega)\exp(i\mathbf{k}\cdot\mathbf{r})$, $\hat{~}$ refers to
Nambu space, and $[\ldots]_{\alpha\beta}$ are Nambu spinor indices.
The $T$-matrix represents the solution to the scattering problem $\hat{T}%
=\hat{V}+\hat{V}\hat{G}^{0}\hat{T}$ for $N$ impurities of identical potential
$\hat{V}_{i}$ at $\mathbf{R}_{i}$.
Note that here we have already made an important approximation, neglecting
Umklapp processes (that is, using one argument in the bare Green's functions
instead of two). We will discuss this approximation in the Appendix A.

It was shown in Refs. \onlinecite{Capriotti03} and \onlinecite{ZAH04} that the
poles of the many-impurity response (change in LDOS) are identical to the
single-impurity problem, although the weights may differ substantially. In
addition, in the limit of weak impurity potentials, the response for the two
problems are identical modulo a disorder-dependent factor\cite{Capriotti03}.
For clarity, we therefore consider the simpler problem and present results for
a single impurity, and replace the full $T$-matrix by $\hat{T}_{ij}%
\rightarrow\hat{t}_{i}\delta_{ij}$, with $\hat{t}_{i}=[1-\hat{V}_{i}\hat
{G}^{0}(\mathbf{r}=0)]^{-1}\hat{V}_{i}$. \vskip.2cm It was shown in Ref.
\onlinecite{Sprunger97} for simple metals, and generalized to unconventional
superconductors by Ref. \onlinecite{Hoffman02}, that information about the
pure electronic system could be extracted from STS measurements by examining
the Fourier transform of dI/dV maps. The Fourier transform of the LDOS is
$\rho(\mathbf{q},\omega)=\sum_{\mathbf{r}\in L\times L}e^{-i\mathbf{q}%
\cdot\mathbf{r}}\rho(\mathbf{r},\omega)$, where $L\times L$ is a square set of
$L^{2}$ positions at which measurements are made, and $\mathbf{q}=2\pi(m,n)/L$
are vectors in the associated reciprocal lattice. The result for a single band
is then
\begin{align}
\delta\rho(\mathbf{q},\omega)  &  =\frac{1}{\pi}\operatorname{Im}%
\sum_{\mathbf{k}}\left[  \hat{G}^{0}(\mathbf{k},\omega)\hat{t}(\omega)\hat
{G}^{0}(\mathbf{k}+\mathbf{q},\omega)\right]  _{11}\label{drhoFourier}\\
&  ={\frac{1}{2}}\mathrm{Tr}\,\operatorname{Im}\sum_{\mathbf{k}}(\tau_{0}%
+\tau_{3})\hat{G}^{0}(\mathbf{k},\omega)\hat{t}(\omega)\hat{G}^{0}%
(\mathbf{k}+\mathbf{q},\omega).\nonumber
\end{align}
Here $\hat{\tau}_{i},~i=0,3$ are the Pauli matrices spanning Nambu space.

\subsection{2 band model: \textbf{q}-integrated LDOS peaks}

Several works have already established the basic generalization of
(\ref{drhoFourier}) to multiband
models\cite{Akbari2010,sykora,Buechner11,wang,Plamadela,zhang09} . The
qualitatively new aspect is that impurities can scatter quasiparticles between
bands, and if the state is of $s_{\pm}$ type, between bands where the
superconducting order parameter $\Delta_{\mathbf{k}}$ has opposite sign. The
various types of scattering processes can then be classified according to
whether they connect portions of the Fermi surface with different gap sign or
not, as was done previously in the
cuprates\cite{WangLee03,Nunner2006,Hanaguri2009}. Without loss of generality
for our qualitative purposes, we consider a pointlike scatterer with Nambu and
band space potential $\hat{V}_{\mu\nu}(\mathbf{k},\mathbf{k}^{\prime}%
)\simeq\sum_{\alpha}V_{\mu\nu}^{\alpha}\tau_{\alpha}$, with band indices
$\mu,\nu=h,e$, where $h$ and $e$ are just band indices, although for typical
Fe pnictides they correspond to hole and electron pockets, and $\tau_{3}$ and
$\tau_{1}$ correspond, respectively, to nonmagnetic and Andreev scattering, which we
discuss separately. Weak, purely magnetic scatterers do not couple to the
spin-averaged local density of states measured by a typical STM experiment, so
we ignore this possibility for the moment (while in Ref. \onlinecite{Hanaguri2009} it was
incorrectly suggested that magnetic scattering contributes to QPI with a
different coherence factor, this mistake was corrected in a later paper by the
same authors\cite{Maltseva2009}). Note that the generalization to 3 or more bands
in the Fe-based systems is straightforward and should not change our basic conclusions.

In the Fe-based systems, the Fermi pockets corresponding to bands 1 and 2 are
well-separated in momentum space and generally have small radius. It is
therefore reasonable to expect that isolated spots of scattering intensity
will be observed generically at small $\mathbf{q}$, corresponding to intraband
scattering processes, and at large $q$, corresponding to
interband\cite{Chi_etal_2014}. The intraband term is a simple sum of the
1-band expression applied to each band separately,
\begin{align}
\delta\rho(\mathbf{q\sim}{0},\omega)  & \label{drhoFourier_intra}\\
={\frac{1}{2}}\mathrm{Tr}\,\operatorname{Im}  &  \sum_{\mathbf{k}\nu}(\tau
_{0}+\tau_{3})\hat{G}_{\nu}^{0}(\mathbf{k},\omega)\hat{t}_{\nu\nu}(\omega
)\hat{G}_{\nu}^{0}(\mathbf{k}+\mathbf{q},\omega).\nonumber
\end{align}
The form of $\hat{t}_{\mu\nu}(\omega)$ is known for simple
cases\cite{Efremov11} but can be a complicated function of the various
integrated Green's function components, so we do not specify it here. Full
expressions are given in the Appendix C.

Suppose now that we wish to calculate the total weight in the small
$\mathbf{q}$ QPI spot as a function of frequency, defined to be
\begin{eqnarray}
\lefteqn{\delta\rho_{intra}(\omega)  = }&& \nonumber \\
& = & {\frac{1}{2}}\mathrm{Tr}\,\operatorname{Im}\sum_{\mathbf{k},\mathbf{q}%
\sim0,\nu}(\tau_{0}+\tau_{3})\hat{G}_{\nu}^{0}(\mathbf{k},\omega)\hat{t}%
_{\nu\nu}(\omega)\hat{G}_{\nu}^{0}(\mathbf{k}+\mathbf{q},\omega) \nonumber \\
&  \approx & {\frac{1}{2}}\mathrm{Tr}\,\operatorname{Im}\sum_{\mathbf{k}%
,\mathbf{q},\nu}(\tau_{0}+\tau_{3})\hat{G}_{\nu}^{0}(\mathbf{k},\omega)\hat
{t}_{\nu\nu}(\omega)\hat{G}_{\nu}^{0}(\mathbf{k}+\mathbf{q},\omega)\nonumber\\
&  = & {\frac{1}{2}}\mathrm{Tr}\,\operatorname{Im}\sum_{\mathbf{k},\mathbf{k}%
^{\prime},\nu}(\tau_{0}+\tau_{3})\hat{G}_{\nu}^{0}(\mathbf{k},\omega)\hat
{t}_{\nu\nu}(\omega)\hat{G}_{\nu}^{0}(\mathbf{k}^{\prime},\omega),
\label{drhoFourier_intra_q_integrated}
\end{eqnarray}
where in the second step we extended the sum over the small range of $\mathbf{q}$
around $\mathbf{q}=0$ to the full Brillouin zone, since the product of two
Green's functions from the same band is sharply peaked at small $\mathbf{q}$
for small $\omega$. In the last step, we expressed the double $\mathbf{k}$ sum
as independent sums over the Nambu Green's functions, which then decouple
provided that $t$ is momentum independent, as we have assumed for the moment.
In the simplest approximation with a flat DOS near the Fermi level, the
integrated matrix Green's function is
\begin{equation}
\sum_{\mathbf{k}}\hat{G}_{\nu}^{0}(\mathbf{k},\omega)\simeq i\pi\rho_{\nu
}{\frac{\omega\tau_{0}+\Delta_{\nu}\tau_{1}}{\sqrt{\omega^{2}-\Delta_{\nu}%
^{2}}}}.
\label{integrated_G}
\end{equation}
Thus one can use Eq. (\ref{integrated_G}) to perform all momentum integrations in
Eq. (\ref{drhoFourier_intra_q_integrated}) and obtain closed form expressions for
the intraband $\mathbf{q}$-integrated LDOS weight $\delta\rho_{intra}(\omega)$
or the corresponding interband quantity describing scattering between two
pockets separated by $\mathbf{q}_{0}$,
\begin{align}
\delta\rho_{inter}(\omega)\equiv{\frac{1}{2}}\mathrm{Tr\operatorname{Im}}%
\sum_{\mathbf{q}\sim\mathbf{q}_{0}}\delta\rho(\mathbf{q},\omega)  &
\label{drhoFourier_inter_q_integrated}\\
={\frac{1}{2}}\mathrm{Tr}\,\operatorname{Im}\sum_{\mathbf{k},\mathbf{k}%
^{\prime},\mu\neq\nu}(\tau_{0}+\tau_{3})\,\hat{G}_{\mu}^{0}(\mathbf{k}%
,\omega)  &  \hat{t}_{\mu\nu}(\omega)\hat{G}_{\nu}^{0}(\mathbf{k}^{\prime
},\omega).\nonumber
\end{align}
Note that we also performed a calculation for the lattice (momentum-resolved) based model with two bands, giving parabolic
like electron and hole band dispersions near the $\Gamma$ and the M points of the Brillouin zone.  The results are shown in \textcolor{blue}{Section III}. Most importantly, the main features, especially the $T$-dependence of the antisymmetric correction to the LDOS and its frequency dependence, allowing one to distinguish $s_{++}$ and $s_{\pm}$ superconducting gaps, continue to hold.

\section{Results: weak potential scatterers}

\subsection{$T=0$ frequency dependence}

\label{subsec:T0freq}To complete the solution, the $t$-matrix for a given
impurity type must be specified. Here we argue that the basic qualitative
features of the QPI patterns that are sensitive to the sign change of the
order parameter (or lack thereof) depend only on the Nambu space structure of
the $t$-matrix, which can be extracted by constructing the components of the
conductance properly symmetrized and antisymmetrized with respect to bias.
They do \textit{not} depend on the detailed energy dependence of the complex
$t$-matrix, except insofar as impurity bound states are created within the
gap. Even in this case, the question of $s_{\pm}$ or $s_{++}$ can be decided
by means described below.

\begin{figure}[ptb]
\renewcommand{\baselinestretch}{.8}
\includegraphics[angle=0,width=1\linewidth]{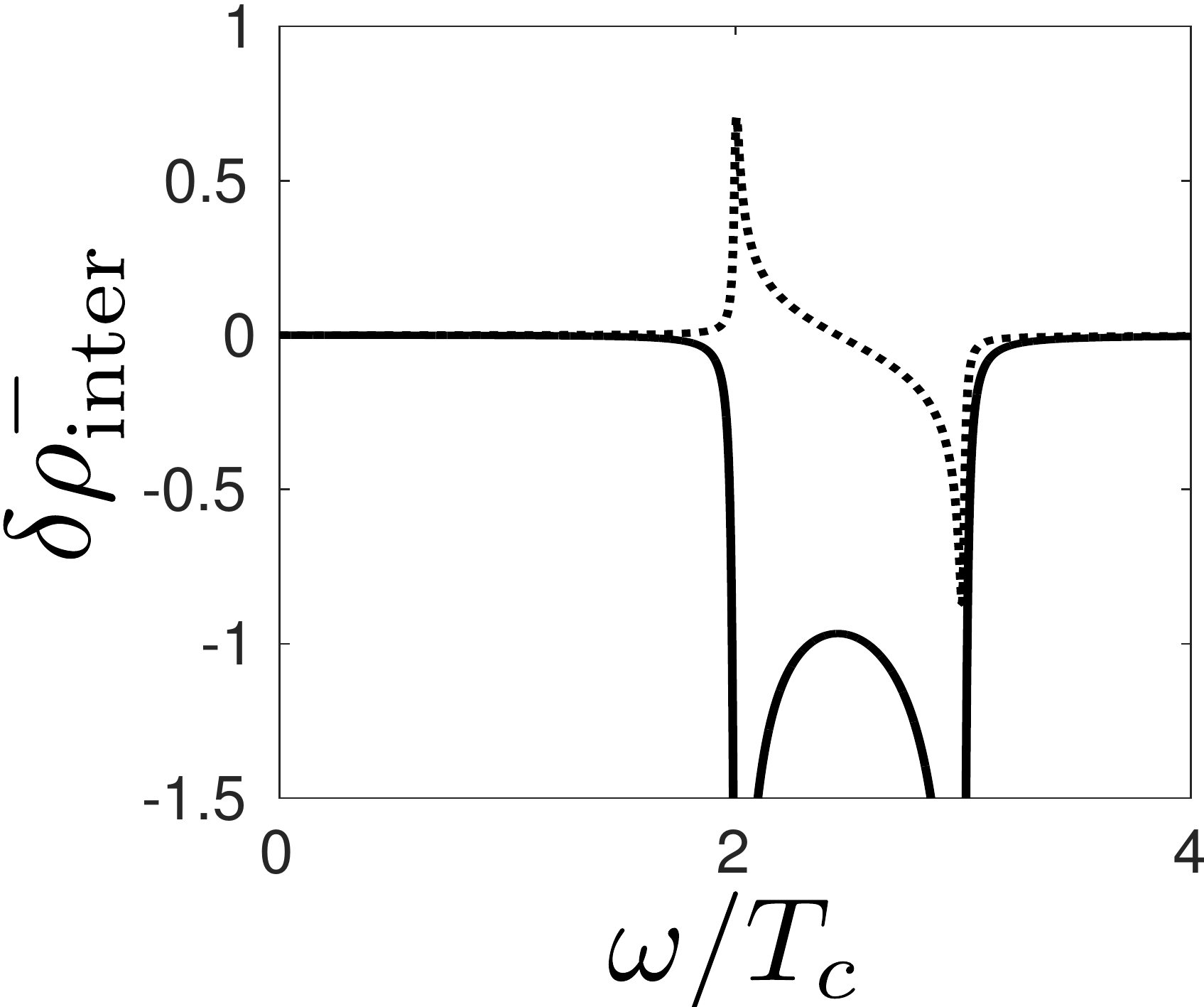}\caption{Integrated
interband density of states $\delta\rho_{inter}/(t_{3}\rho_{1}\rho_{2})$ for
constant (weak) $\tau_{3}$ $t$-matrix. Gap magnitudes are $|\Delta_{1}%
|/T_{c}=3$, $|\Delta_{2}|/T_{c}=2$, and artificial broadening $\eta
=10^{-3}$. Solid line: $s_{\pm}$ state; Dashed: $s_{++}$. All other components
of $\delta\rho$ are zero for this case. }%
\label{fig:omega_dep}%
\end{figure}

We show this by first considering the case of constant $t$-matrix, valid for
weak (Born) impurity scattering. While in general $\hat{t}$ has several Nambu
components depending on the type of scattering, superconducting state and
impurity phase shift, it is instructive to focus on one Nambu component at a
time. For example, if $\hat{t}=t_{3}\tau_{3}$, as \textit{e.g.} for a weak
nonmagnetic scatterer, then%

\begin{align}
\delta\rho_{intra}(\omega)  &  \approx\label{eq:intra_tau3}\\
-{\frac{\pi^{2}}{2}}t_{3}\sum_{\nu}\rho_{\nu}^{2}  &  \mathrm{Im}%
{\frac{\mathrm{Tr}(\tau_{0}+\tau_{3})(\omega\tau_{0}+\Delta_{\nu}\tau_{1}%
)\tau_{3}(\omega\tau_{0}+\Delta_{\nu}\tau_{1})}{\omega^{2}-\Delta_{\nu}^{2}}%
}\nonumber\\
&  =0,\nonumber
\end{align}
in other words, within this approximation, there is \textit{no change} in the
small-$\mathbf{q}$ integrated Fourier transform density of states due to a
nonmagnetic intraband scatterer, regardless of the relative sign of the two gaps
$\Delta_{h,e}$. The same is not true in general of the interband contribution
$\delta\rho_{inter}(\omega)\equiv{\frac{1}{2}}\mathrm{Tr}\mathrm{Im}%
\sum_{\mathbf{q}\sim\mathbf{q}_{0}}\delta\rho(\mathbf{q},\omega)$, where
$\mathbf{q}_{0}$ is the wave vector connecting the two Fermi surface pockets.
In this case we have, again for the $\tau_{3}$ component of the $t$-matrix,%

\begin{align}
\delta\rho_{inter}(\omega)  &  \approx\label{eq:rho_inter}\\
-2{\pi^{2}}t_{3}\rho_{h}\rho_{e}  &  \mathrm{Im}{\frac{\mathrm{Tr}(\tau
_{0}+\tau_{3})(\omega\tau_{0}+\Delta_{h}\tau_{1})\tau_{3}(\omega\tau
_{0}+\Delta_{e}\tau_{1})}{\sqrt{\omega^{2}-\Delta_{h}^{2}}\sqrt{\omega
^{2}-\Delta_{e}^{2}}}}\nonumber\\
&  =-2{\pi^{2}}t_{3}\rho_{h}\rho_{e}\mathrm{Im}{\frac{\omega^{2}-\Delta
_{h}\Delta_{e}}{\sqrt{\omega^{2}-\Delta_{h}^{2}}\sqrt{\omega^{2}-\Delta
_{e}^{2}}}},\nonumber
\end{align}
which is manifestly nonzero for $|\Delta_{e}|<\omega<|\Delta_{h}|$.
Furthermore it is easy to show that in the limit when the two gap magnitudes
become equal, $|\Delta_{e}|\rightarrow|\Delta_{h}|$, there are two distinct
cases. For an $s_{\pm}$ state, $\operatorname{sgn}\Delta_{h}\Delta_{e}<0$,
$\delta\rho_{inter}(\omega)=-2\pi^{2}\rho_{1}\rho_{2}\Delta_{h}\delta
(\omega-|\Delta_{h}|)$, while for an $s_{++}$ state, $\delta\rho
_{inter}\rightarrow0$. In the more general case with $\Delta_{h}\neq\Delta
_{e}$, the interband response in the $s_{\pm}$ case remains generically much
larger than that in the $s_{++}$ case, with weight concentrated between the
two energies $\omega=|\Delta_{h}|,|\Delta_{e}|$. In Fig. \ref{fig:omega_dep},
we plot the interband frequency-dependent $\mathbf{q}$-integrated LDOS change
for a model constant $\tau_{3}$ $t$-matrix to illustrate this difference. Note
not only the change in sign of the $s_{++}$ response due to the numerator of
(\ref{eq:rho_inter}) which vanishes at an energy corresponding to the the
geometric mean of the two gaps in this case, but also the overall small scale.

 For a realistic scatterer, the Born limit results given above are no longer
adequate, and the $t$-matrix aquires components in all Nambu channels. It is
instructive to consider the response of the system to a scatterer with
constant $t$-matrix in each of these channels, even if none of these cases
corresponds to a physical situation with the exception of the $\tau_{3}$ (weak
potential) scatterer\footnote{In a $d$-wave superconductor, the large impurity
potential limit yields $\hat{t}\propto\tau_{0}$, but in an $s$-wave system the
nonvanishing integrated anomalous Green's functions generate $\tau_{1}$ and
$\tau_{2}$ terms as well.}.  Thus one can define
the experimentally accessible quantities
\[
\rho_{%
\genfrac{}{}{0pt}{}{intra}{inter}%
}^{(\pm)}(\omega)=\rho_{%
\genfrac{}{}{0pt}{}{intra}{inter}%
}(\omega)\pm\rho_{%
\genfrac{}{}{0pt}{}{intra}{inter}%
}(-\omega),
\]
and use them to make clear qualitative predictions for the existence or
nonexistence of a strong QPI response in the various channels that are
independent of the type of scatterer, as summarized in Table
\ref{table:drho_freq}. These features will also correspond to peaklike
features in the $T$ dependence of the integrated QPI intensity, as we discuss below.\\
 In particular, it is easy to see that the $\tau_{0}$
component of the $t$-matrix reverses the responses of $s_{++}$ and $s_{\pm}$
compared to the $\tau_{3}$ scatterer.
For the intraband part, we get immediately $\delta\rho_{intra}(\omega
)=-\pi^{2}t_{0}\sum_{\nu}\rho_{\nu}^{2}\Delta_{\nu}\delta(\omega-|\Delta_{\nu
}|)$ for both $s_{\pm}$ or $s_{++}$. The interband processes contribute $0$
for an $s_{\pm}$ state and $-2\pi^{2}\rho_{h}\rho_{e}\Delta_{h}\delta
(\omega-\Delta_{h})$ for an $s_{++}$ state. { Finally, for a single impurity
with constant $t$-matrix in the $\tau_{1}$ channel, one easily finds that the
intraband symmetric densities of states $\delta\rho_{intra}^{+}$ are singular.
The antisymmetric interband density of states $\delta\rho_{inter}^{-}$
vanishes identically, but the symmetric part $\delta\rho_{inter}^{+}$ can be
large. The weight is proportional to $\omega(\Delta_{h}+\Delta_{e})$, so
taking our limit $|\Delta_{h}|\rightarrow|\Delta_{e}|$ for qualitative
comparison as above, we see that the $s_{++}$ case is singular while the
$s_{\pm}$ case vanishes. }

Although a generic nonmagnetic impurity has all Nambu components of the
$t$-matrix, which will mix these behaviors, they can be isolated {to some
extent} by constructing the symmetrized and antisymmetrized densities of
states, respectively, as suggested by Maltseva and
Coleman\cite{Hanaguri2009,Maltseva2009}. This is because the $\tau_{3}$
component of the $t$-matrix generates only odd frequency changes to the
density of states to all orders in perturbation theory, and the $\tau_{0}$ and
$\tau_{1}$ component only even ones. The $\tau_{2}$ component does not
contribute to the change in the Fourier transformed LDOS to all orders in perturbation theory simply because of the Trace over Nambu matrices, together with the assumption that the gaps are real (i.e. appear with $\tau_1$). Furthermore, this statement is correct even if the particle-hole symmetry is broken (i.e. the normal part of the Nambu Green's functions has additional $\tau_3$ component).
\begin{table}[th]%
\begin{tabular}
[c]{|c|c|c|c|}\hline
&  & \textit{intra} & \textit{inter}\\\hline
$s_{++}$ & $\delta\rho^{(+)}$ & $\tau_{0},\tau_{1}$ & $\tau_{0},\tau_{1}$\\
& $\delta\rho^{(-)}$ & $\times$ & $\times$\\\hline
$s_{\pm}$ & $\delta\rho^{(+)}$ & $\tau_{0},\tau_{1}$ & $\times$\\
& $\delta\rho^{(-)}$ & $\times$ & $\tau_{3}$\\\hline
\end{tabular}
\caption{Possibility of singular integrated QPI intensity (Fourier transformed
density of states) in the symmetric (+) and antisymmetric (-) channels for
$s_{++}$ and $s_{\pm}$ superconductors. Here $\tau_{\alpha}$ indicates the
presence of a strongly enhanced response for an assumed constant $t$-matrix in
the $\alpha$ Nambu channel, and the $\times$ indicates the absence of one.
Magnetic impurities have Nambu symmetry $\tau_{0}\sigma_{z}$ and do not
modulate the total LDOS within the Born approximation (see text).}%
\label{table:drho_freq}%
\end{table}

\subsection{Thermal average: STM observable}

\begin{figure}[ptb]
\renewcommand{\baselinestretch}{.8}
\includegraphics[angle=0,width=.8\linewidth]{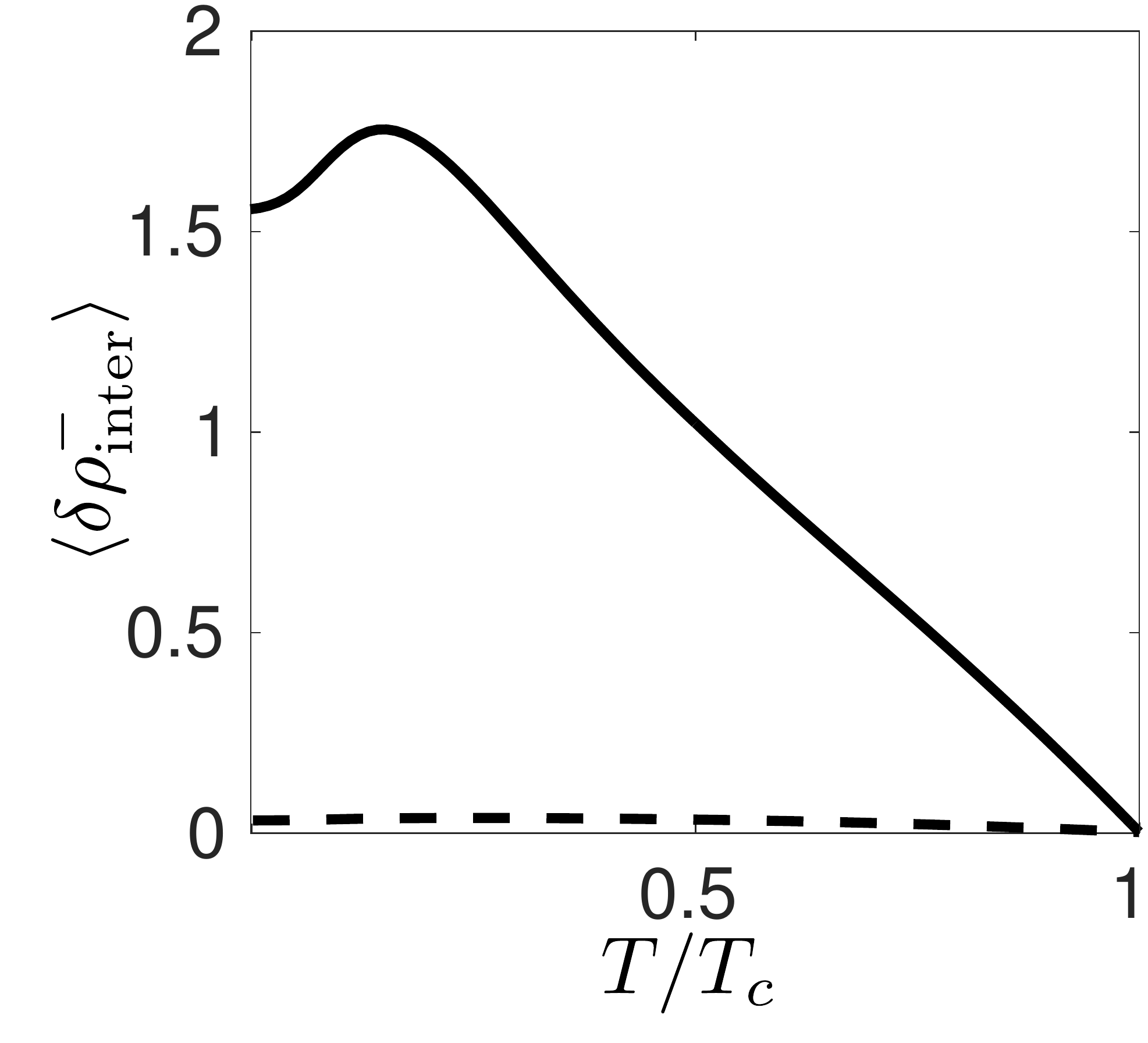}
\caption{Thermally averaged interband $\mathbf{q}$-integrated LDOS change
within constant density of states and constant $t$-matrix approximation for
weak nonmagnetic scatterer with $\tau_{3}$ (Born limit) potential. The
external frequency $\Omega$ was taken to be 2.5 in units where $|\Delta_{1}|$
and $|\Delta_{2}|$ were 3.0 and 2.0, as in Fig. \ref{fig:omega_dep}. Shown is
antisymmetrized LDOS $\langle\delta\rho^{-}(\Omega)\rangle$ for $s_{++}$
(dashed) and for $s_{\pm}$ state (solid). Symmetrized components
are zero for both states. }%
\label{fig:drho_avg}
\end{figure}

Our intention is to make clear qualitative predictions for observable
quantities in STM experiments. Thus far, we have shown only the $T=0$ results
for the artificial case of a constant real $t$-matrix. In any measurement at
finite temperature, the conductance will be related to the change in LDOS
$\delta\rho$ convolved with a thermal factor weighting the contribution of
different electronic states to the current. The conventional
result\cite{Hoffman_rev}, translated into our notation, is%

\begin{eqnarray}
\lefteqn{\langle\delta\rho^{\pm} (\Omega)\rangle \equiv} &&\\
&&  \int_{-\infty}^{\infty}d\omega\delta\rho^{\pm}(\omega)\left[
{\frac{-\partial f}{\partial\omega}} (\omega+\Omega)\pm{\frac{ -\partial
f}{\partial\omega}} (\omega-\Omega) \right] \nonumber
\end{eqnarray}
In Fig. \ref{fig:drho_avg}, we plot the nonzero antisymmetrized $\mathbf{q}$-integrated density of states for the interband
QPI peak in both $s_{++}$ and $s_{\pm}$ states in this simple approximation.
Measurement of the antisymmetric components of the interband
density of states alone should suffice to qualitatively distinguish the two
states. Note that the clearest results are obtained when the STM bias corresponds
to an energy between the two gap energies, which can be identified from local
tunneling spectra.

It appears at first glance from Table \ref{table:drho_freq} that measurement
of the symmetrized interband $\mathbf{q}$-integrated LDOS might also
distinguish the two states. However, it is important to stress here that there
is no physical impurity in the case of an $s$-wave superconductor
corresponding to a situation where the $t$-matrix is entirely of $\tau_{0}$
type; other components of the $t$-matrix are mixed at strong impurity
potentials. This we show below for the realistic Coulomb screened impurity
potential. In particular, we will show that very little difference between
$s_{++}$ and $s_{\pm}$ will be observed in the symmetrized channel for
realistic situations. We therefore believe that the symmetrized channel
should be ignored in the analysis of STM data.

It is useful to observe that the temperature dependent average conductances
predicted here bear no relation to the standard forms one might expect were
quasiparticle interference really described by conventional coherence factors,
as anticipated in Refs. \onlinecite{Hanaguri2009},\onlinecite{Maltseva2009}.
This issue is discussed in some length in Appendix B.

\section{General scattering potentials}

\subsection{Realistic screened Coulomb potentials}

The $t$-matrix for a single impurity is given by
\begin{equation}
\underline{{t}}=\left[  1-\underline{{U}}\tau_{3}\sum_{\mathbf{k}%
}{\underline{{G}}(\mathbf{k},\omega)}\right]  ^{-1}{\tau_{3}}\underline{{{U}}}%
\end{equation}
where the integrated matrix Green's function given in Eq.(\ref{integrated_G})
can be further re-written as $\left(  \sum_{\mathbf{k}}\hat
G(\mathbf{k},\omega)\right)  _{\nu\nu}=g_{\omega,\nu}\tau_{0}+g_{\Delta,\nu
}\tau_{1}$ for each of the bands. The scattering potential matrix \underline{$U$} can be
then separated into intraband $\hat{U}_{aa}=\hat{U}_{bb}%
=U_{intra}\tau_{0}$ and the interband $\hat{U}_{ab}=U_{inter}\tau_{0}$ term.

\begin{figure}[ptb]
\renewcommand{\baselinestretch}{.8}
\includegraphics[angle=0,width=\linewidth]{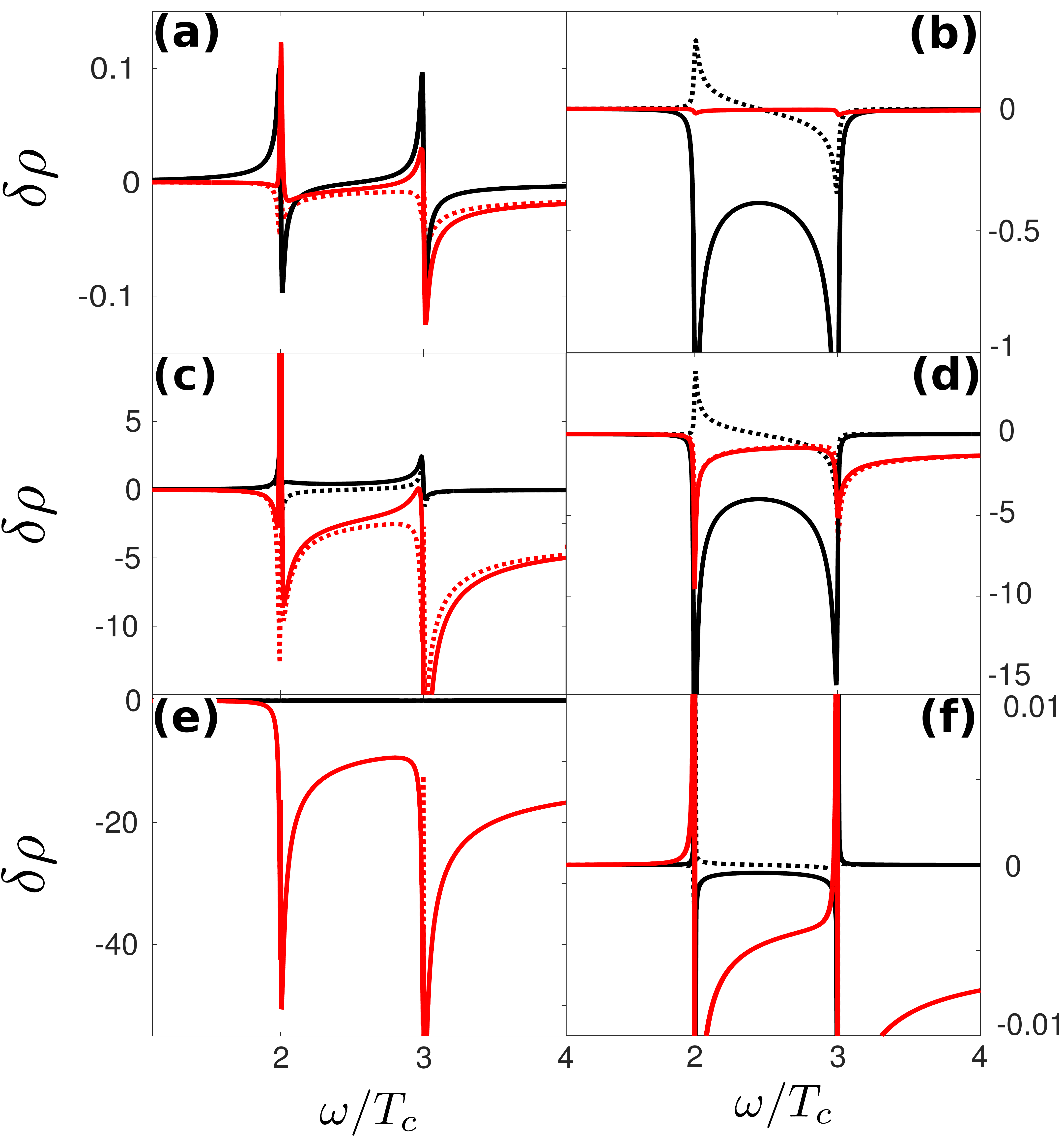}\caption{Integrated
density of states components $\delta\rho_{intra,inter}^{\pm}$ for
\textquotedblleft realistic" screened Coulomb scatterers $U_{inter}=0.2U_{intra}$, for
$U_{intra}=0.01$ (a,b), $0.2$ (c,d), and $10.$ (e,f). Dashed curves correspond to
$s_{++}$ state, solid curves to $s_{\pm}$. Red and black are even and odd
components of the LDOS, $\delta\rho^{+}$ and $\delta\rho^{-}$ , respectively.
Left panels represent intraband, right panels interband scattering channels,
respectively.}%
\label{fig:u_ne_v}%
\end{figure}

We have shown that a constant $t$-matrix in the Born limit leads to a clear
prediction of QPI intensities enabling one to distinguish $s_{++}$ from
$s_{\pm}$ states. For intermediate to strong scatterers, however, the $t$
matrix is complex and frequency dependent, and includes $\tau_{1}$ and
$\tau_{2}$ components in addition to $\tau_{0}$ and $\tau_{3}$. One may thus
be concerned that our conclusions may not be general, given that one does not
know \textit{a priori} the strength of impurities giving rise to the QPI
signal. We therefore present results for the full $t$-matrix of a single
impurity of arbitrary strength and intra- vs. interband scattering potential,
within the flat normal state DOS approximation (Eq. \ref{integrated_G}). Since
for large $q$ a screened Coulomb potential falls off like $1/\left(q_{TF}^{2}+q^{2}\right)$, and screening lengths $2\pi/q_{TF}$ are of order the unit cell size,
realistic interband scattering potentials are smaller than intraband
potentials. We therefore begin in Fig. \ref{fig:u_ne_v} by fixing
$U_{inter}=0.2\,U_{intra}$ for various strengths of $U_{intra}$.
To analyze these results, it is useful to identify exactly what
terms distinguish $s_{++}$ and $s_{\pm}$ states. In the simple pedagogical
example with which we began, there were no such terms in the intraband
scattering channel, Eq. (\ref{eq:intra_tau3}), and one term in the
antisymmetrized interband channel expression for $\delta\rho_{inter}^{-}$ Eq.
(\ref{eq:rho_inter}) proportional to $\Delta_{h}\Delta_{e}$. The full result
in the general scattering potential case is given in the Appendix C, but we find
analogously that the intraband LDOS has no proportionality to terms sensitive
to sign changes. In the interband channel there are no $\Delta_{h}\Delta_{e}$ terms in
the symmetrized LDOS, whereas the antisymmetrized LDOS $\delta\rho_{inter}%
^{-}$ still contains only a single term proportional to
\begin{equation}
U_{1}\mbox{Im}{\frac{\Delta_{h}\Delta_{e}}{\sqrt{\omega^{2}-\Delta_{h}^{2}}\sqrt
{\omega^{2}-\Delta_{e}^{2}}}},
\end{equation}
\textit{i.e.} precisely the same expression as in the simpler example of the Born limit. All
terms will be multiplied by the denominator of the full $t$-matrix, of course,
which also contains terms that weakly distinguish $s_{++}$ and $s_{\pm}$
states but this does not alter our qualitative conclusions.

Nevertheless, we find that the singular behavior of $\delta\rho_{inter}^{-} $
in the $s_{\pm}$ case is preserved until the unitary limit is reached. In
section \ref{subsec:fullTdependence} below, we exhibit the experimentally
observable consequences of this effect.

\subsection{Role of bound states}

\begin{figure}[ptb]
\renewcommand{\baselinestretch}{.8}
\includegraphics[angle=0,width=\linewidth]{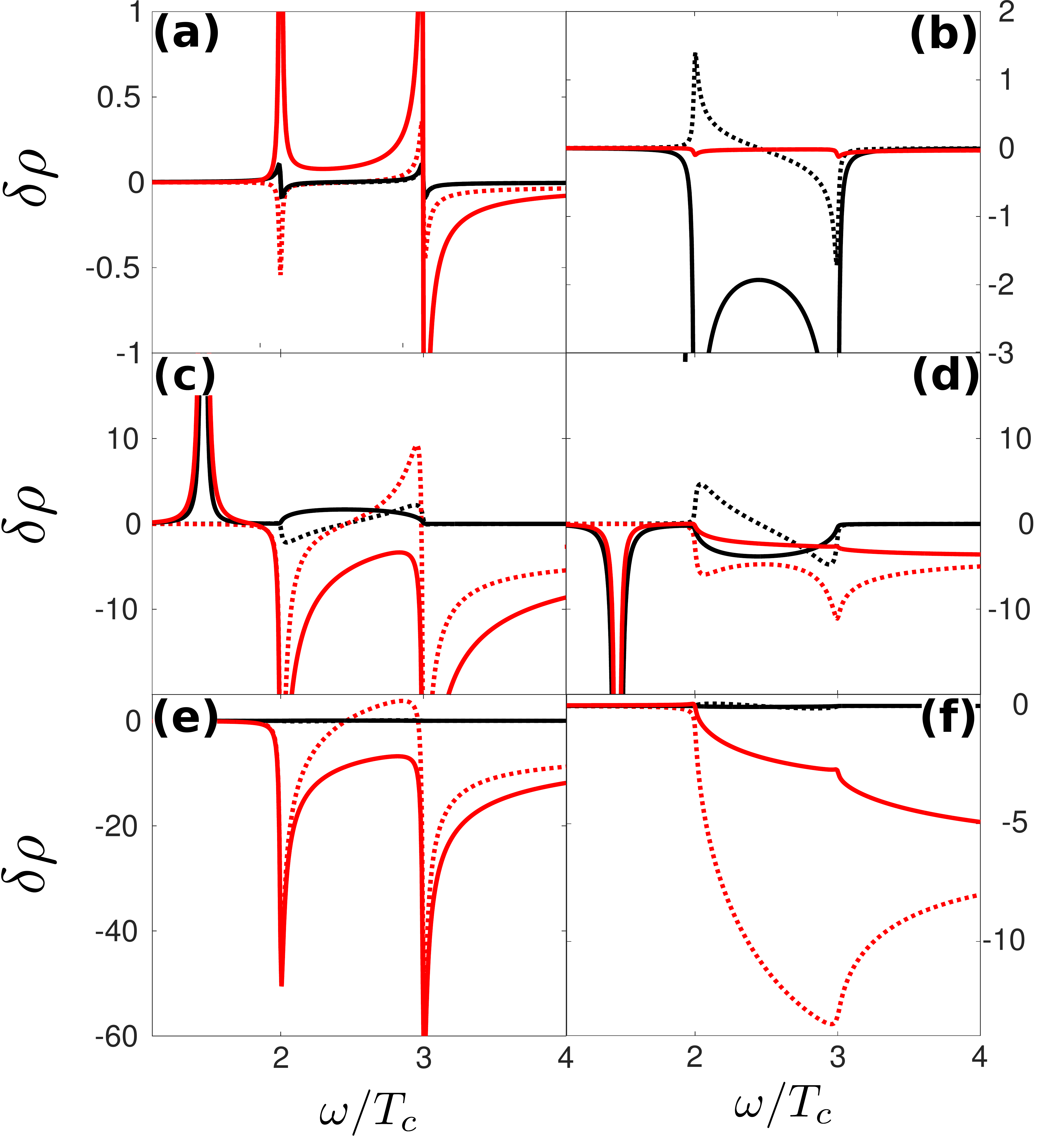}
\caption{Integrated
density of states components $\delta\rho_{intra,inter}^{\pm}$ for isotropic
scatterers $U_{intra}=U_{inter}$ for $U_{intra}=0.01$ (a,b), $0.2$ (c,d), and $10.$ (e,f).
Dashed curves correspond to $s_{++}$ state, solid curves to $s_{\pm}$. Red and
black are even and odd components of the LDOS, $\delta\rho^{+}$ and
$\delta\rho^{-}$ , respectively. Left panels represent intraband, right panels
interband scattering channels, respectively.}%
\label{fig:u_eq_v}%
\end{figure}

In the cases discussed in Fig. \ref{fig:omega_dep}, no impurity bound states
are visible in the subgap region below $|\Delta_{e}|=2$. In general, the
formation of impurity bound states in a multiband system is more complicated
than in a 1-band superconductor, and it has been argued that such states are
indeed nongeneric, requiring fine-tuning of the potential to produce a bound
state below the lower gap edge\cite{HKM_ROPP11,Beaird2012}. In the 2-band
model, subgap bound states are found only in a very narrow interval around a
line $U_{inter}(U_{intra})$ in impurity potential space which approaches
$U_{inter}=U_{intra}$ for strong impurities\cite{HKM_ROPP11}. We therefore discuss the case
$U_{inter}=U_{intra}$ separately here.

Figures \ref{fig:u_eq_v}(a) and \ref{fig:u_eq_v}(b) essentially reproduce the results of Fig.
\ref{fig:u_ne_v} for a weak scatterer. With increased scattering strength,
however, bound states are formed in the $s_{\pm}$ case, as seen in Fig.
\ref{fig:u_eq_v}(d). If the bound state is at low energy, it steals so much
weight from the coherence peak LDOS that the hierarchy of intensities
represented in Table \ref{table:drho_freq} becomes a bit difficult to
distinguish. To use QPI as a definitive qualitative tool, it may therefore be
necessary to consider only systems without impurity bound states in the gap,
which occurs sometimes, or to mask the impurities that give rise to bound
states in the spatial window used for the Fourier transform. Of course, the
observation of bound states in and of itself is strong evidence for $s_{\pm}$
pairing, provided the nonmagnetic character of the impurities can be reliably
assumed. If no bound states occur, the hierarchy of LDOS moments of Table
\ref{table:drho_freq} is clear (compare Fig. \ref{fig:u_ne_v}(c,d)).

In the unitary limit, Fig. \ref{fig:u_eq_v} (e-f) the bound state has already
moved through the gap and the basic hierarchy is again preserved. Note that
for this simple band, this limit is essentially achieved already for
potentials of order $T_{c}$.

\subsection{Finite temperatures}

\label{subsec:fullTdependence} 
\begin{figure}[ptb]
\renewcommand{\baselinestretch}{.8}
\includegraphics[angle=0,width=1.0\linewidth]{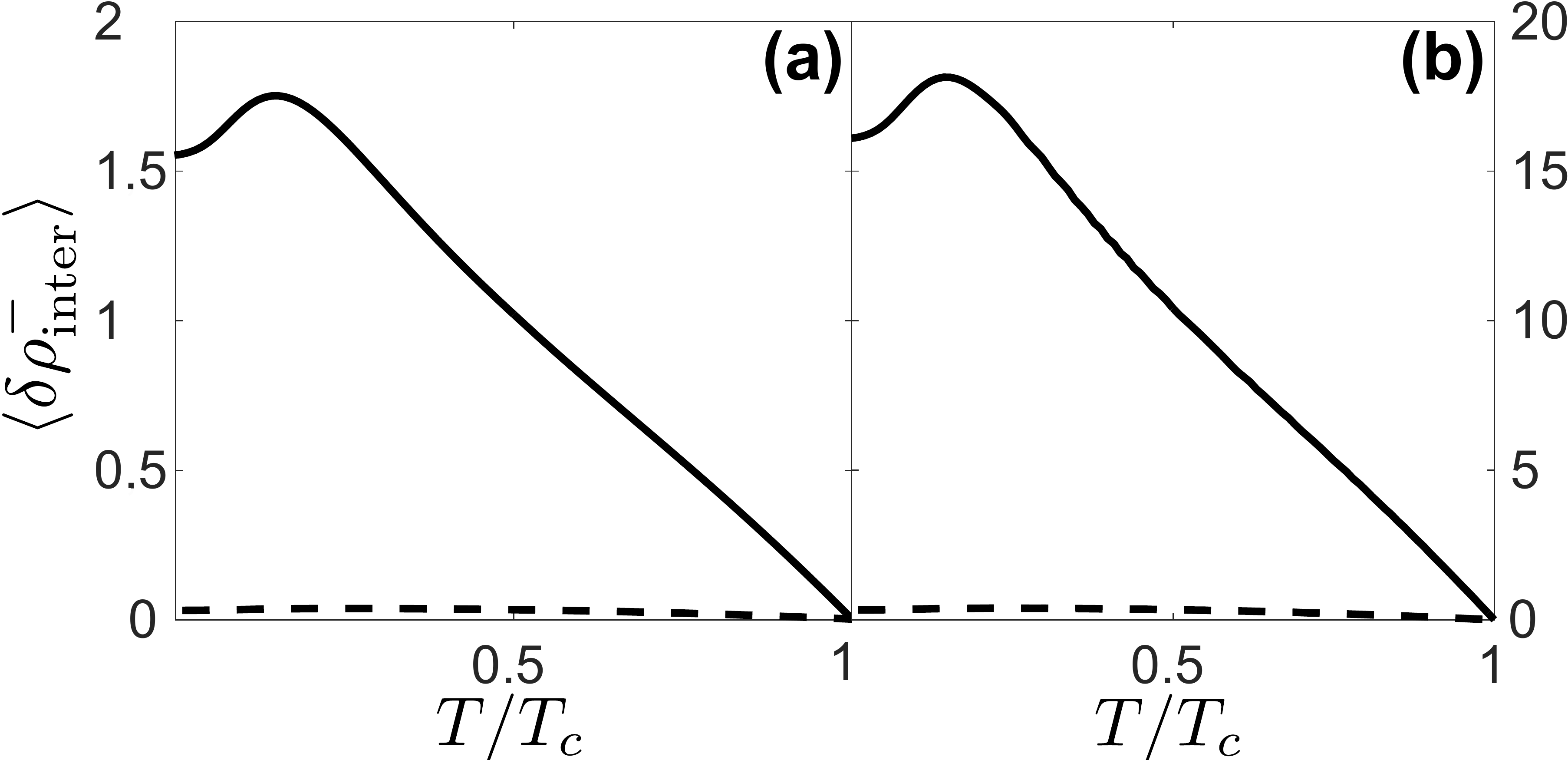}
\caption{Thermally averaged antisymmetrized interband $\mathbf{q}$-integrated LDOS change
$\langle\delta\rho_{inter}^{-}\rangle(\Omega)$ for nonmagnetic
scatterers with parameters $U_{inter}=0.2U_{intra}$, with full $t$-matrix. The
external frequency $\Omega/T_{c}=2.5$, with $|\Delta_{1}|/T_{c}$ and
$|\Delta_{2}|/T_{c}$ taken as 3.0 and 2.0, respectively. (a) refers to the weak potential
$U_{intra}=0.01$. Shown are antisymmetrized LDOS $\langle\delta\rho^{-}%
(\Omega)\rangle$ for $s_{++}$ (dashed) and for $s_{\pm}$ state (solid).
(b) shows the same for intermediate strength potential $U_{intra}=0.2$. Here we have used the standard BCS type behavior for both superconducting gaps. }%
\label{fig:finiteT}%
\end{figure}

For transparency, we now remove from
consideration those components of the $\mathbf{q}$-integrated LDOS which are
not qualitatively affected by a gap sign change, and plot in Fig.
\ref{fig:finiteT} the thermally averaged, antisymmetrized, interband integrated
LDOS for the two states $s_{++}$ and $s_{\pm}$, Eq.(8). Here we have used the standard BCS type $T$-dependence for both superconducting gaps.
Thermal averaging has the effect of removing
many of the sharp spectral features at the gap edge, as discussed
pedagogically for the case of a constant $t$-matrix in Fig. \ref{fig:drho_avg}
above. It is clear that the temperature dependent signal in the $s_{\pm}$ case
is huge and characteristic, whereas in the $s_{++}$ it is small and
featureless.
Note that the slight decrease of $\langle\delta\rho_{inter}^{-}\rangle$
occurs when the width of the peaks in the $\delta\rho_{inter}^{-}$ becomes
comparable to the thermal broadening of the derivative of the Fermi function.
It is also clear that the $T$-dependencies do \textit{not} resemble the classic
BCS $T$-dependencies arising from coherence factors in, \textit{e.g.} NMR
relaxation and acoustic attenuation cases (see the Appendix B). We propose that
the measurement of $\langle\delta\rho_{inter}^{-}\rangle$ versus temperature
is therefore the clearest way of identifying a gap sign change using QPI.

\subsection{Comparison with experiment}

\begin{figure}[ptb]
\renewcommand{\baselinestretch}{.8}
\includegraphics[angle=0,width=\linewidth]{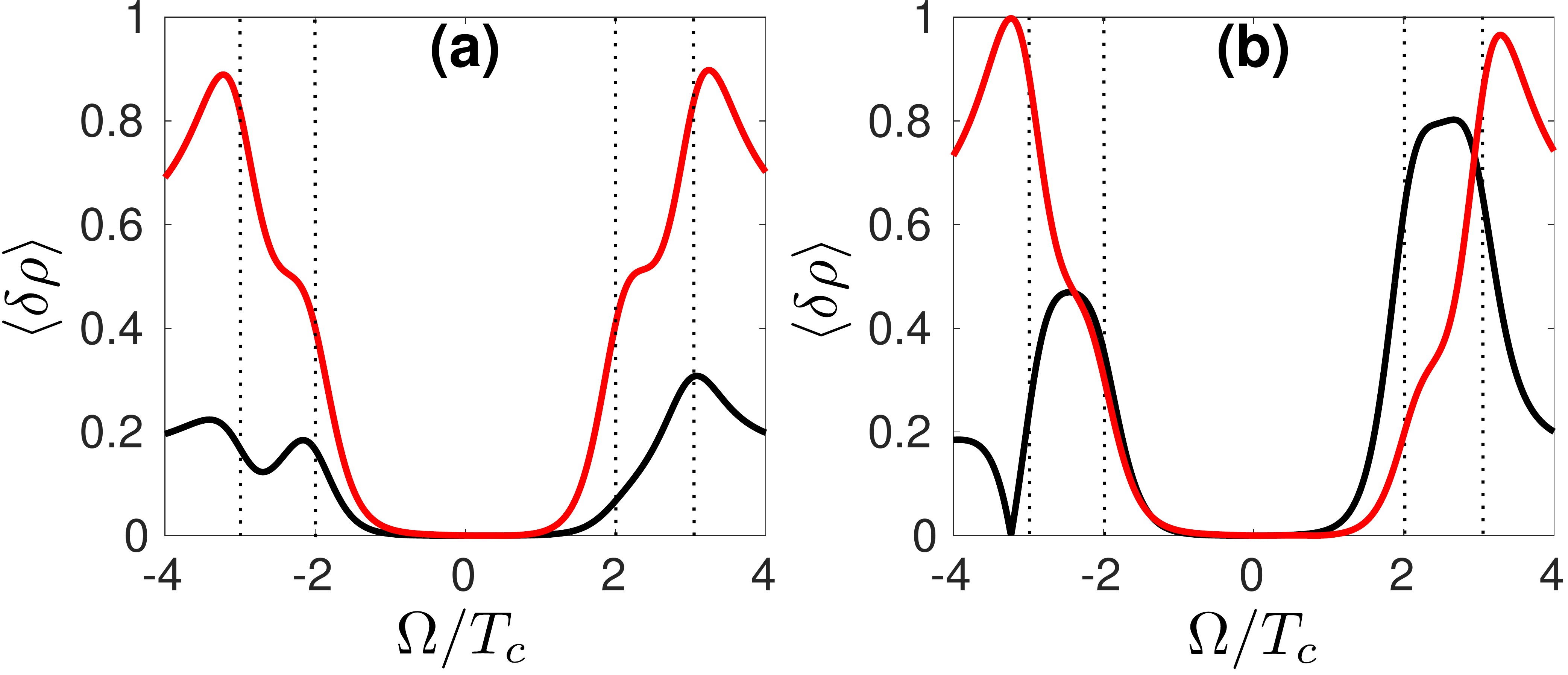}
\caption{Absolute magnitude of thermally averaged \textit{unsymmetrized}
interband (black) and intraband (red) $\mathbf{q}$-integrated LDOS changes
$|\langle\delta\rho_{inter,intra}\rangle(\Omega)|$ for  $s_{++}$ (a) and $s_{\pm}$ (b) vs. $\Omega/T_{c}$ for
nonmagnetic scatterers with parameters $T=0.2T_{c}$ and $U_{inter}=0.2U_{intra}$, with
full $t$-matrix, and $|\Delta_{1}|/T_{c}$ and $|\Delta_{2}|/T_{c}$ taken as
3.0 and 2.0, respectively. $\Omega=|\Delta_{1}|$ and $|\Delta_{2}|$ are
indicated by dashed lines.}%
\label{fig:unsymm}%
\end{figure}

We have shown in the previous section that the temperature dependence of the
antisymmetrized integrated LDOS is a sensitive measure of the gap sign change
in a superconductor, although it does not reduce in any limit to the BCS-type
temperature dependence expected from coherence factors. Here we comment on the
results of Chi \textit{et al}. \cite{Chi_etal_2014} on LiFeAs, who neither
symmetrized their integrated conductance maps, nor made use of the sign of the
LDOS change, but nevertheless obtained what appeared to be a distinct
qualitative result. Shown in Fig. \ref{fig:unsymm} are the expected
$|\langle\delta\rho_{inter,intra}\rangle(\Omega)|$ for both $s_{\pm}$ and
$s_{++}$ states within our model, with the two gap scales indicated. While the
overall behavior is somewhat complicated, one can see that it is indeed the
case, as inferred by the authors of Ref. \onlinecite{Chi_etal_2014}, that the
bias dependence for an $s_{+-}$ state immediately below the the upper gap scale $|\Delta_{h}|$ is
opposite for the intra- and interband contributions, while it is the same in
the case of an $s_{++}$ state. Note further that had one not known the gap
scales exactly, distinguishing between the two states on the basis of this
type of measurement might have been difficult: the relative magnitudes of
intra-and interband contributions are probably not to be taken seriously,
since it is difficult to subtract the $\mathbf{q}$ weight of the homogeneous
system from the intraband. The gap scales are presumed to be known from STM
experiment from direct measurements of the coherence peaks in the real space
local conductance $\propto\langle\delta\rho\rangle(\mathbf{r},\Omega)$, but
they will be shifted somewhat from the underlying values by thermal smearing
and gap anisotropy. The former type of shift is even evident in Fig.
\ref{fig:unsymm}. In case of more than two distinct gap values the results
are even more muddled. These caveats are among the reasons why we propose that
a $T$-dependent measurement at fixed frequency, ideally between the two gap
scales, should be a more sensitive measure of the gap sign change.

\begin{figure}[ptb]
\renewcommand{\baselinestretch}{.8}
\includegraphics[angle=0,width=\linewidth]{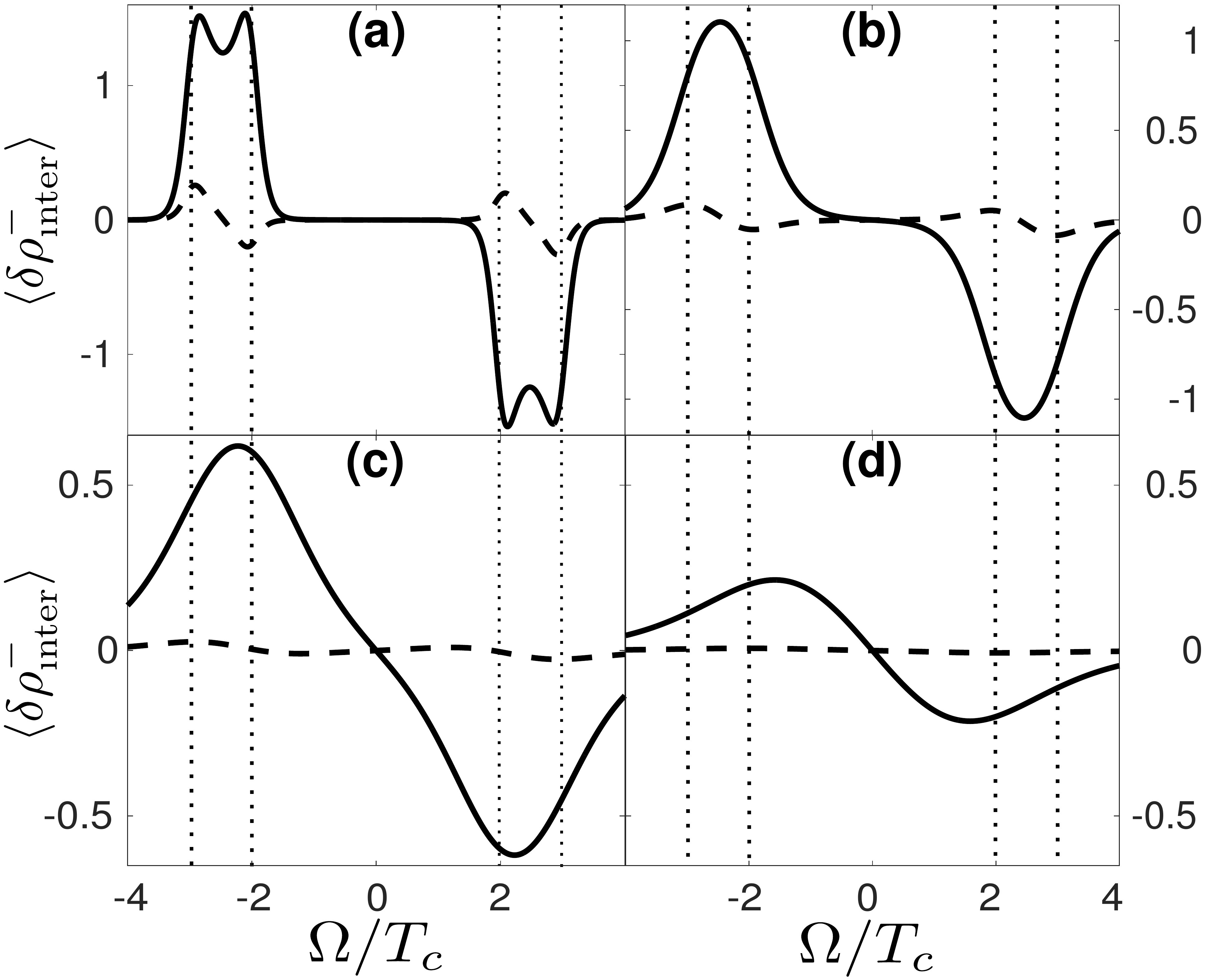}\caption{Thermally
averaged LDOS changes for antisymmetric, interband channel, $|\langle
\delta\rho_{inter,intra} \rangle(\Omega)| $ vs. $\Omega/T_{c}$. Curves shown
are for nonmagnetic scatterers with parameters $T/T_{c}=0.1$ (a), 0.3 (b), 0.6 (c), 0.9 (d) and
$U_{inter}=0.2 U_{intra}$, with full $t$-matrix, and $|\Delta_{1}|/T_{c}$ and
$|\Delta_{2}|/T_{c}$ taken as 3.0 and 2.0, respectively. $\Omega=|\Delta_{1}|$
and $|\Delta_{2}|$ are indicated by dashed lines. Solid curve: $s_{\pm}$. Dashed
curve: $s_{++}$.}%
\label{fig:rho_odd_fixedT_vs_w}%
\end{figure}

To illustrate what one should expect, we now plot in Fig.
\ref{fig:rho_odd_fixedT_vs_w} the antisymmetric component of the interband
LDOS as a function of bais $\omega$, for different temperature. This is the quantity that should show the
most prominent difference between $s_{++}$ and $s_{\pm}$. In particular, this
component for the $s_{++}$ case should exhibit a sign change at a frequency
corresponding to the geometric mean of the two gap scales, while $s_{\pm}$ has
a finite large value there.  With further increase of the temperature, the
main features decrease in amplitude but should remain detectable. We expect
that this will be the typical behavior seen in experiment.

\begin{figure}[ptb]
\renewcommand{\baselinestretch}{.8}
\includegraphics[angle=0,width=1.0\linewidth]{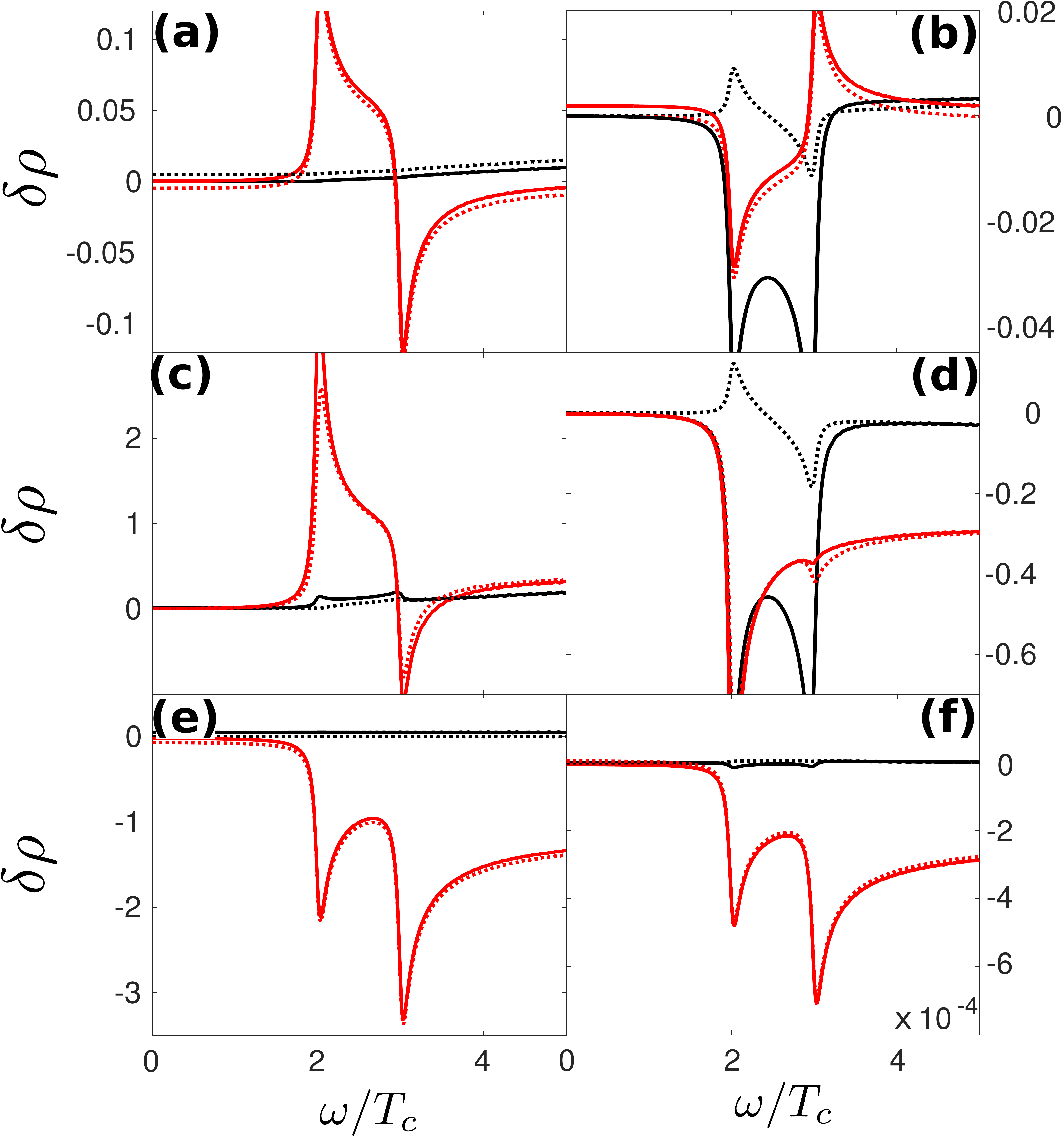}
\caption{Integrated
density of states components $\delta\rho_{intra,inter}^{\pm}$ for isotropic
scatterers $U_{inter}=0.2U_{intra}$ for $U_{intra}=0.01$ (a,b), $0.2$ (c,d), and $10$ (e,f) (all in units of energy as in for the momentum independent model), computed for momentum resolved Green's functions, Eq.(25). As in Fig.3, the dashed curves correspond to $s_{++}$ state, solid curves to $s_{\pm}$. Red and
black are even and odd components of the LDOS, $\delta\rho^{+}$ and
$\delta\rho^{-}$ , respectively. Left panels represent intraband, right panels
interband scattering channels, respectively.}%
\label{fig:appnedixd1}%
\end{figure}

\begin{figure}[ptb]
\renewcommand{\baselinestretch}{.8}
\includegraphics[angle=0,width=1.0\linewidth]{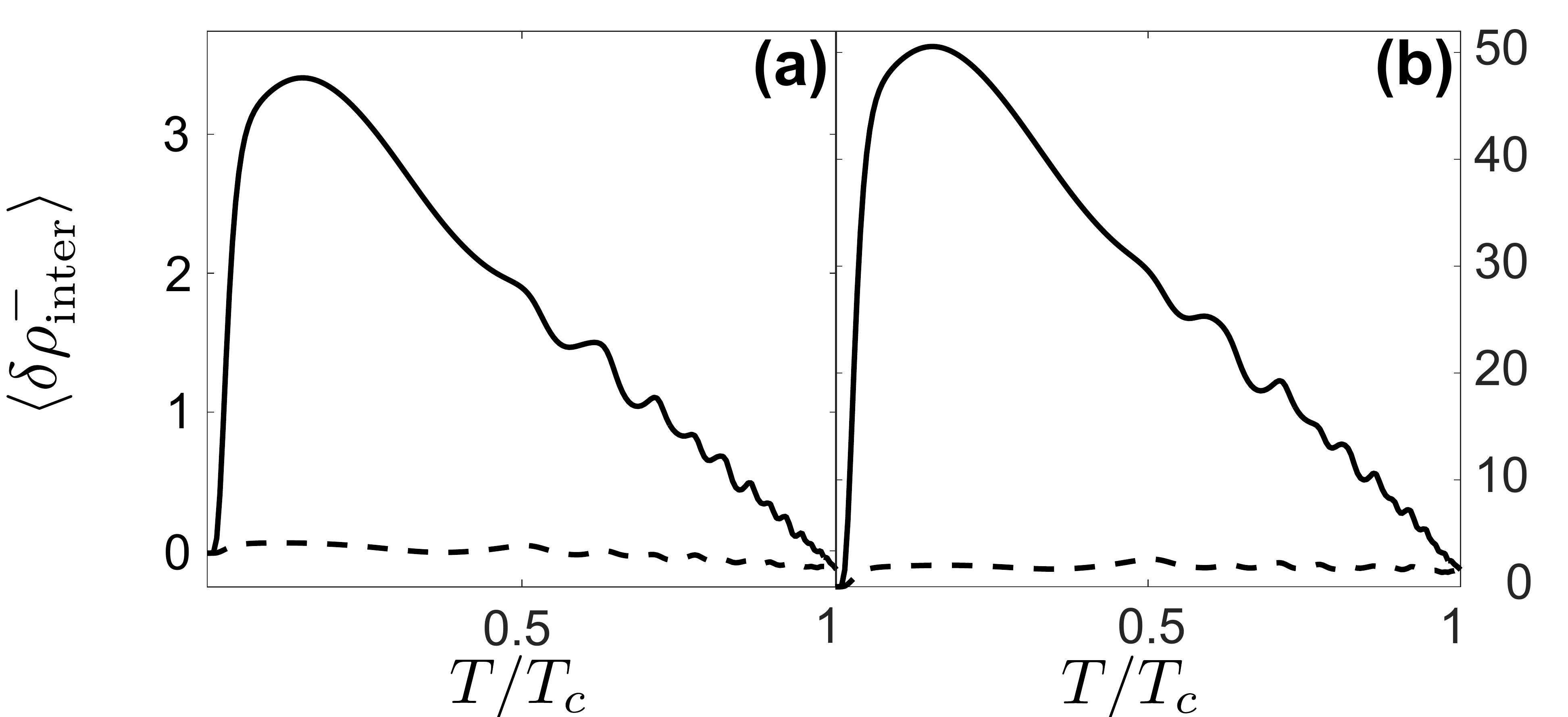}
\caption{Thermally averaged antisymmetrized interband $\mathbf{q}$-integrated LDOS change
$\langle\delta\rho_{inter}^{-}\rangle(\Omega)$  for nonmagnetic
scatterers with parameters $U_{inter}=0.2U_{intra}$, with full $t$-matrix and momentum resolved Green's function, Eq(25). The
external frequency $\Omega/T_{c}=2.5$, with $|\Delta_{1}|/T_{c}$ and
$|\Delta_{2}|/T_{c}$ taken as 3.0 and 2.0, respectively. (a) refers to the intermediate potential
$U_{intra}=0.2$. Shown are antisymmetrized LDOS $\langle\delta\rho^{-}%
(\Omega)\rangle$ for $s_{++}$ (dashed) and for $s_{\pm}$ state (solid).
(b) shows the same for the larger strength potential $U_{intra}=10$. As in Fig.5, here we have used the standard BCS type behavior for both superconducting gaps.}%
\label{fig:appnedixd2}%
\end{figure}

\subsection{Effect of particle-hole asymmetry: momentum-resolved Green's functions}
\begin{figure}[ptb]
\renewcommand{\baselinestretch}{.8}
\includegraphics[angle=0,width=1.0\linewidth]{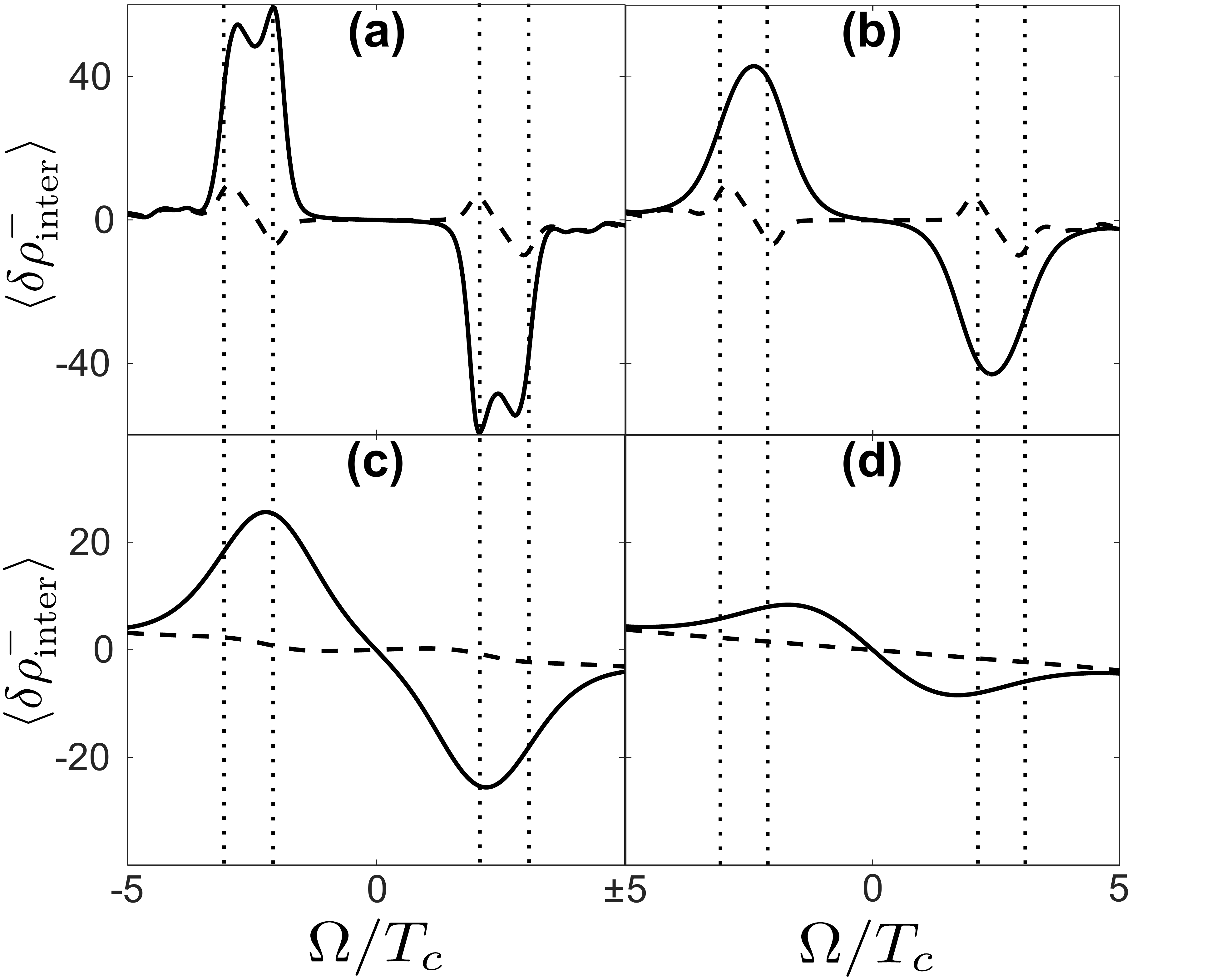}
\caption{Thermally
averaged LDOS changes for antisymmetric, interband channel, $|\langle
\delta\rho_{inter,intra} \rangle(\Omega)| $ vs. $\Omega/T_{c}$. Curves shown
are for nonmagnetic scatterers with parameters $T/T_{c}=0.1$ (a), 0.3 (b), 0.6 (c), 0.9 (d) and
$U_{inter}=0.2 U_{intra}$ ($U_{intra}=0.2$), with full $t$-matrix, momentum-resolved Green's functions, Eq.(25), and $|\Delta_{1}|/T_{c}$ and
$|\Delta_{2}|/T_{c}$ taken as 3.0 and 2.0, respectively. $\Omega=|\Delta_{1}|$
and $|\Delta_{2}|$ are indicated by dashed lines. Solid curve: $s_{\pm}$. Dashed
curve: $s_{++}$.}%
\label{fig:appnedixd3}%
\end{figure}

To complete our analysis, we also performed a calculation assuming a lattice based model with two bands, giving parabolic like electron and hole band dispersions near the $\Gamma$ and the $M$ points of the Brillouin zone. In particular, we consider the simplest band topology of the iron-based superconductors with hole band, $\varepsilon_h ({\bf k})=\mu_h-\frac{{\bf k}^2}{2m_h}$, centered near the $\Gamma$-point of the Brillouin zone and the electron band, $\varepsilon_e ({\bf k})=\frac{{\bf (k+Q)}^2}{2m_e}-\mu_e$,  centered near the $M$ point of the Brillouin Zone ({\bf Q}$=(\pi,\pi)$). Here, we set $\frac{1}{2m_h}=\frac{1}{2m_e}=34$ and $\mu_e=-10.6$, $\mu_h=9.4$
(all in the same units of energy as in the main text). The most important part introduced by the lattice based models is the electron hole asymmetry in the normal part of the Nambu Green's functions,
\begin{equation}
\hat{G}_{\nu}^{0}(\mathbf{k}, i \omega_n)\simeq - \frac{i\omega_n\tau_{0}+\varepsilon_{\nu}({\bf k})\tau_{3}+\Delta_{\nu}\tau_{1}}{\omega_n^{2}+\Delta^2_{\nu}+\varepsilon^2_{\nu}({\bf k})}.
\label{G_k}
\end{equation}
Substituting the Green's function for the electron and hole bands, Eq.(\ref{G_k}) into Eqs.(3),(5) and (9) we computed the interband and inraband corrections to the LDOS similar to the main text. In Figs.\ref{fig:appnedixd1}-\ref{fig:appnedixd3} we show the corresponding results for the integrated density of states components $\delta\rho_{intra,inter}^{\pm}$,  thermally averaged antisymmetrized interband $\mathbf{q}$-integrated LDOS change
$\langle\delta\rho_{inter}^{-}\rangle(\Omega)$, and thermally
averaged LDOS changes for antisymmetric, interband channel, $|\langle
\delta\rho_{inter,intra} \rangle(\Omega)| $ vs. $\Omega/T_{c}$, respectively. Despite some differences, the main features that
allow one to distinguish $s_{++}$ and $s_{\pm}$ superconducting gaps continue to hold also if one uses more realistic Green's functions.
We believe this happens because the electron and hole Fermi surfaces and corresponding gaps are well separated in ${\bf q}$ space.  This allows one to disentangle intra- and inter-band scattering processes in the normal and superconducting states clearly, as found in most of the ferropnictides. In particular,  we observe that the actual $T/T_c$ dependence for the $s_{\pm}$-wave symmetry shows a non-monotonic dependence which is reflected in the peak-like structure at small $T/T_c$ values and  corresponding downward behavior at low $T/T_c$ values. This is absent for the $s_{++}$-wave symmetry. In addition, we see from Fig. \ref{fig:appnedixd3} that the antisymmetrized STM signal for $|\Delta_1|\lesssim\Omega\lesssim|\Delta_2|$ has one sign for $s_\pm$ and changes sign for $s_{++}$. Therefore we believe that these two key features can be observed in the experiments.

\subsection{Remarks on magnetic and $\tau_{1}$ (``vortex") scatterers}

A weak \textquotedblleft magnetic impurity" represented by an isolated classical spin
that couples via exchange to conduction electron spin density may be written
in Nambu space as $V_{spin}\tau_{0}\sigma_{3}$, where $\sigma_{3}$ is the
Pauli matrix in spin space. Within the Born approximation, the change to the
up-spin LDOS will cancel that of the down-spin LDOS. Higher order magnetic
scatterings will produce effects, but since in general a chemical substituent
with a magnetic moment will have a nonmagnetic scattering potential much
larger than its magnetic one, we ignore this effect here. Including transverse
spin couplings (or deep $d$- or $f$-levels within an Anderson model approach)
will result in Kondo physics which obviously produces interesting effects on
the density of states, including Kondo resonances near the Fermi level, with
concomitant influence on QPI; these have been recently discussed
elsewhere\cite{Derry15}. Similar effects on the QPI  would be expected when the Yu-Shiba bound state is induced by the magnetic
impurity. In order to draw qualitative conclusions regarding
gap symmetry using the methods described here, samples or regions of samples
displaying Kondo and/or Yu-Shiba bound state resonances should be excluded.

Scattering in the $\tau_{1}$, or Andreev channel has been discussed in several
contexts in the field of unconventional superconductivity. Chemical impurities
suppress the order parameter in their vicinity, creating an effective
off-diagonal local potential which contributes to the scattering of
quasiparticles\cite{HettlerHirschfeld1999,Shnirman1999}. Normally these
effects are ignored, e.g. in standard $t$ matrix calculations, or treated as
weak, but under some circumstances they can become important. If an impurity
has the effect of enhancing the pairing locally, as occurs in some
models\cite{Buzdin87,Nunner2005,Maska2007,Foyevtsova09,Foyevtsova10}, the
$\tau_{1}$ potential component of an impurity can be significant and even
control the behavior of the conductance spectrum and map\cite{Nunner2005}.

The order parameter is also suppressed near vortex cores, and Pereg-Barnea and
Franz suggested that this fact could be used to provide a method of
controlling disorder and distinguishing gap symmetries \textit{in situ},
provided the vortex lattice were sufficiently disordered\cite{PeregBarnea2008}%
. Here we do not discuss $\tau_{1}$ chemical impurities in detail, as we are
focussed primarily on qualitative aspects of QPI, but we discuss the
oft-repeated statement that the effect of the disordered vortex lattice,
represented by a random, tunable set of $\tau_{1}$ scatterers can distinguish
sign-preserving and sign-reversing QPI peaks.
It is believed \cite{Hanaguri2009,Hanaguri2010,Chi_etal_2014} that the peaks
whose weight is enhanced in a field correspond to sign-preserving peaks, while
those whose weight is suppressed by a field are sign-reversing.

In fact a clear statement to this effect is difficult to make.{ As we showed
in Sec. \ref{subsec:T0freq}, $\tau_{1}$ impurities indeed enhance the QPI
signal in the intraband channel, which represents sign-preserving scattering
for both superconducting states considered. In addition, Table
\ref{table:drho_freq} also shows that the sign-preserving interband
scatterings in the $s_{++}$ case give rise to an enhancement. There is no
indication of a \textit{suppression} with field in the sign-reversing ($s$%
}$_{{\pm}}${ interband) case, however. } This apparent discrepancy was noted
already by Pereg-Barnea and Franz\cite{PeregBarnea2008}, who suggested that
the disordered vortex lattice led to a random phase potential experienced by
quasiparticles, which might give rise to an overall suppression of the QPI
signal. Such a \textquotedblleft background" suppression could then be
overcompensated by the singular enhancements of the sign-preserving scattering
wave vectors. There is, however, no calculation to support this assertion, so
statements about determining gap symmetry from QPI peaks in unconventional
superconductors from their field behavior should be treated with caution (see
also the discussion in the Appendix).

\subsection{Application to other states; nodeless $d$-wave}

While for pedagogical reasons we have restricted ourselves to two bands,
isotropic gaps, and either $s_{++}$ or $s_{\pm}$ states, the concepts we have
discussed are clearly applicable to more general situations. The interband
entries of Table \ref{table:drho_freq} labelled $s_{++}$ apply generally to
gap sign preserving transitions between bands, and those labelled $s_{\pm}$
apply to sign-changing ones. The obvious example under discussion in the
Fe-based superconductivity field is the putative $d$ wave state in systems
with no hole pockets but four electron pockets at the $(\pm\pi, 0)$ and
$(0,\pm\pi)$, intitially proposed for the alkali-intercalated FeSe
materials\cite{FWang_dwave_2011,Graser_dwave_2011}. In addition to Bragg
scattering and small $q$ intraband scattering, interband scattering should
give rise to rings of scattering intensity around $\mathbf{q}=(0,\pm2\pi)$ and
$(\pm2\pi,0)$, as well as around $(\pm2\pi,\pm2 \pi)$. Predictions for
intraband scattering weights will be identical to those listed for $s_{\pm}$
in Table \ref{table:drho_freq}.

\section{Conclusions}

In this paper we have argued that the task of identifying order parameter
symmetries in unconventional superconductors via QPI measurements is unlikely
to be successful if it relies, as in the past, on comparisons of theoretical
conductance maps with experiment. This is because there are too many unknown
parameters, particularly in multiband systems, to allow for a quantitative
theory of QPI. It is possible that this situation can be improved to some
extent by \textit{ab initio} based calculation of the density of states away
from the metal surface, including an isolated impurity. This would be however
time consuming and would still not eliminate all quantitative
assumptions. On the other hand, in systems like the FeSC, where one can
clearly identify QPI peaks related to intra- and interband scattering, the
temperature dependence of the integrated weights of these peaks can provide a
robust qualitative means of detecting order parameter sign changes that can be
used to determine its structure.

We have focussed our attention on a model of a 2-band system with two distinct
gaps $\Delta$, and shown that the most sensitive way to distinguish scattering
processes connecting gaps of same or different sign is to operate with STM
bias in the energy region between the two gap scales, identification of which is a relatively simple experimental task. We
have then shown that the $T$-dependent response of the symmetrized and
anitsymmetrized combinations of the conductance for both intra- and interband
scattering provide a characteristic signature of a gap sign change or lack
thereof. These temperature dependencies do \textit{not}, as suggested by
previous works, correspond to thermal averages of simple BCS coherence
factors, but are somewhat more complicated. In particular, we find that the
effect is strongest at low temperatures, and not near $T_{c}$, in contrast to
the expectation assuming coherence factors. Although we have focussed on the
question of distinguishing $s_{\pm}$ and $s_{++}$ states in the FeSC, it is
clear that similar arguments can be made for sign-changing gaps in other
contexts, for instance a putative $d$-wave state in FeSC materials with
electron pockets only.

In the past, most QPI experiments have focussed on the power spectrum of the
LDOS, in other words the absolute magnitude of the density of states Fourier
transform $|\delta\rho(\mathbf{q})|$ or related ratios of this quantity,
so-called $Z$ or $R$ maps. We have shown here that measurement of the
\textit{signed} symmetrized and antisymmetrized QPI maps are crucial to
extract symmetry information in the absence of detailed knowledge of the
impurity potentials, which is usually the case. This effect persists up to
$T_{c}$.

The existence of order parameter bound states is one aspect which must be
treated with care in such a measurement, as we have shown that they tend to
steal weight from the spectral region where the characteristic distinctions
are most visible. Of course if one has a clear indication of a bound state
induced by a nonmagnetic impurity, it is already a strong indication of a
sign-changing order parameter. Nevertheless additional complementary evidence
can be obtained by performing the analysis suggested here while masking the
regions containing the bound states before Fourier transforming.

Finally, we have discussed the commonly used method of distinguishing gap
symmetries by observing the magnetic field dependence of QPI peaks and
identifying sign-preserving scattering wave vectors as those corresponding to
peaks that increase with field, and sign-changing ones with those that
decrease with field. This analysis, while appealing and possibly correct in
some cases, is based on a questionable analogy of vortices as pointlike
Andreev scatterers that may fail for several reasons, including if the vortex
lattice is too ordered or coherence lengths too large. In addition, while we
can understand within the work presented here why some QPI peaks can be
enhanced by the magnetic field, there is no firm theoretical ground for
interpretation of those peaks that are suppressed.

\textit{Acknowledgements.} We thank P. Coleman, D.J. Scalapino and I. Vekhter
for useful discussions. P.J.H. was supported by NSF-DMR-1005625, and I.I.M. by
the U.S. Office of Naval Research through the Naval Research Laboratory's
Basic Research Program. The work of DA and IE was supported by the Focus Program 1458 Eisen-Pniktide of the DFG, and by the German Academic
Exchange Service (DAAD PPP USA no. 57051534). P.J.H. and I.I.M. would like to thank for hospitality
R. Valenti and the Goethe University of Frankfurt, where this project was
started, as well as the Kavli Institute for Theoretical Physics, where several
key discussions took place. I.E. acknowledges the support allocated to Kazan Federal University for the project part of
the state assignment in the sphere of scientific activities.

\bibliographystyle{apsrev4-1}
\bibliography{spm_qpi_v9}

\begin{thebibliography}{40}%
\makeatletter
\providecommand \@ifxundefined [1]{%
 \@ifx{#1\undefined}
}%
\providecommand \@ifnum [1]{%
 \ifnum #1\expandafter \@firstoftwo
 \else \expandafter \@secondoftwo
 \fi
}%
\providecommand \@ifx [1]{%
 \ifx #1\expandafter \@firstoftwo
 \else \expandafter \@secondoftwo
 \fi
}%
\providecommand \natexlab [1]{#1}%
\providecommand \enquote  [1]{``#1''}%
\providecommand \bibnamefont  [1]{#1}%
\providecommand \bibfnamefont [1]{#1}%
\providecommand \citenamefont [1]{#1}%
\providecommand \href@noop [0]{\@secondoftwo}%
\providecommand \href [0]{\begingroup \@sanitize@url \@href}%
\providecommand \@href[1]{\@@startlink{#1}\@@href}%
\providecommand \@@href[1]{\endgroup#1\@@endlink}%
\providecommand \@sanitize@url [0]{\catcode `\\12\catcode `\$12\catcode
  `\&12\catcode `\#12\catcode `\^12\catcode `\_12\catcode `\%12\relax}%
\providecommand \@@startlink[1]{}%
\providecommand \@@endlink[0]{}%
\providecommand \url  [0]{\begingroup\@sanitize@url \@url }%
\providecommand \@url [1]{\endgroup\@href {#1}{\urlprefix }}%
\providecommand \urlprefix  [0]{URL }%
\providecommand \Eprint [0]{\href }%
\providecommand \doibase [0]{http://dx.doi.org/}%
\providecommand \selectlanguage [0]{\@gobble}%
\providecommand \bibinfo  [0]{\@secondoftwo}%
\providecommand \bibfield  [0]{\@secondoftwo}%
\providecommand \translation [1]{[#1]}%
\providecommand \BibitemOpen [0]{}%
\providecommand \bibitemStop [0]{}%
\providecommand \bibitemNoStop [0]{.\EOS\space}%
\providecommand \EOS [0]{\spacefactor3000\relax}%
\providecommand \BibitemShut  [1]{\csname bibitem#1\endcsname}%
\let\auto@bib@innerbib\@empty
\bibitem [{\citenamefont {Hirschfeld}\ \emph {et~al.}(2011)\citenamefont
  {Hirschfeld}, \citenamefont {Korshunov},\ and\ \citenamefont
  {Mazin}}]{HKM_ROPP11}%
  \BibitemOpen
  \bibfield  {author} {\bibinfo {author} {\bibfnamefont {P.~J.}\ \bibnamefont
  {Hirschfeld}}, \bibinfo {author} {\bibfnamefont {M.~M.}\ \bibnamefont
  {Korshunov}}, \ and\ \bibinfo {author} {\bibfnamefont {I.~I.}\ \bibnamefont
  {Mazin}},\ }\href {http://stacks.iop.org/0034-4885/74/i=12/a=124508}
  {\bibfield  {journal} {\bibinfo  {journal} {Reports on Progress in Physics}\
  }\textbf {\bibinfo {volume} {74}},\ \bibinfo {pages} {124508} (\bibinfo
  {year} {2011})}\BibitemShut {NoStop}%
\bibitem [{\citenamefont {Mazin}\ \emph {et~al.}(2008)\citenamefont {Mazin},
  \citenamefont {Singh}, \citenamefont {Johannes},\ and\ \citenamefont
  {Du}}]{Mazin08}%
  \BibitemOpen
  \bibfield  {author} {\bibinfo {author} {\bibfnamefont {I.~I.}\ \bibnamefont
  {Mazin}}, \bibinfo {author} {\bibfnamefont {D.~J.}\ \bibnamefont {Singh}},
  \bibinfo {author} {\bibfnamefont {M.~D.}\ \bibnamefont {Johannes}}, \ and\
  \bibinfo {author} {\bibfnamefont {M.~H.}\ \bibnamefont {Du}},\ }\href
  {\doibase 10.1103/PhysRevLett.101.057003} {\bibfield  {journal} {\bibinfo
  {journal} {Phys. Rev. Lett.}\ }\textbf {\bibinfo {volume} {101}},\ \bibinfo
  {pages} {057003} (\bibinfo {year} {2008})}\BibitemShut {NoStop}%
\bibitem [{\citenamefont {Kontani}\ and\ \citenamefont
  {Onari}(2010)}]{Kontani10}%
  \BibitemOpen
  \bibfield  {author} {\bibinfo {author} {\bibfnamefont {H.}~\bibnamefont
  {Kontani}}\ and\ \bibinfo {author} {\bibfnamefont {S.}~\bibnamefont
  {Onari}},\ }\href {\doibase 10.1103/PhysRevLett.104.157001} {\bibfield
  {journal} {\bibinfo  {journal} {Phys. Rev. Lett.}\ }\textbf {\bibinfo
  {volume} {104}},\ \bibinfo {pages} {157001} (\bibinfo {year}
  {2010})}\BibitemShut {NoStop}%
\bibitem [{\citenamefont {Maiti}\ \emph {et~al.}(2011)\citenamefont {Maiti},
  \citenamefont {Korshunov}, \citenamefont {Maier}, \citenamefont
  {Hirschfeld},\ and\ \citenamefont {Chubukov}}]{Maiti2011}%
  \BibitemOpen
  \bibfield  {author} {\bibinfo {author} {\bibfnamefont {S.}~\bibnamefont
  {Maiti}}, \bibinfo {author} {\bibfnamefont {M.~M.}\ \bibnamefont
  {Korshunov}}, \bibinfo {author} {\bibfnamefont {T.~A.}\ \bibnamefont
  {Maier}}, \bibinfo {author} {\bibfnamefont {P.~J.}\ \bibnamefont
  {Hirschfeld}}, \ and\ \bibinfo {author} {\bibfnamefont {A.~V.}\ \bibnamefont
  {Chubukov}},\ }\href {\doibase 10.1103/PhysRevLett.107.147002} {\bibfield
  {journal} {\bibinfo  {journal} {Phys. Rev. Lett.}\ }\textbf {\bibinfo
  {volume} {107}},\ \bibinfo {pages} {147002} (\bibinfo {year}
  {2011})}\BibitemShut {NoStop}%
\bibitem [{\citenamefont {Nunner}\ \emph {et~al.}(2006)\citenamefont {Nunner},
  \citenamefont {Chen}, \citenamefont {Andersen}, \citenamefont {Melikyan},\
  and\ \citenamefont {Hirschfeld}}]{Nunner2006}%
  \BibitemOpen
  \bibfield  {author} {\bibinfo {author} {\bibfnamefont {T.~S.}\ \bibnamefont
  {Nunner}}, \bibinfo {author} {\bibfnamefont {W.}~\bibnamefont {Chen}},
  \bibinfo {author} {\bibfnamefont {B.~M.}\ \bibnamefont {Andersen}}, \bibinfo
  {author} {\bibfnamefont {A.}~\bibnamefont {Melikyan}}, \ and\ \bibinfo
  {author} {\bibfnamefont {P.~J.}\ \bibnamefont {Hirschfeld}},\ }\href
  {\doibase 10.1103/PhysRevB.73.104511} {\bibfield  {journal} {\bibinfo
  {journal} {Phys. Rev. B}\ }\textbf {\bibinfo {volume} {73}},\ \bibinfo
  {pages} {104511} (\bibinfo {year} {2006})}\BibitemShut {NoStop}%
\bibitem [{\citenamefont {Pereg-Barnea}\ and\ \citenamefont
  {Franz}(2008)}]{PeregBarnea2008}%
  \BibitemOpen
  \bibfield  {author} {\bibinfo {author} {\bibfnamefont {T.}~\bibnamefont
  {Pereg-Barnea}}\ and\ \bibinfo {author} {\bibfnamefont {M.}~\bibnamefont
  {Franz}},\ }\href {\doibase 10.1103/PhysRevB.78.020509} {\bibfield  {journal}
  {\bibinfo  {journal} {Phys. Rev. B}\ }\textbf {\bibinfo {volume} {78}},\
  \bibinfo {pages} {020509} (\bibinfo {year} {2008})}\BibitemShut {NoStop}%
\bibitem [{\citenamefont {Hanaguri}\ \emph {et~al.}(2009)\citenamefont
  {Hanaguri}, \citenamefont {Kohsaka}, \citenamefont {Ono}, \citenamefont
  {Maltseva}, \citenamefont {Coleman}, \citenamefont {Yamada}, \citenamefont
  {Azuma}, \citenamefont {Takano}, \citenamefont {Ohishi},\ and\ \citenamefont
  {Takagi}}]{Hanaguri2009}%
  \BibitemOpen
  \bibfield  {author} {\bibinfo {author} {\bibfnamefont {T.}~\bibnamefont
  {Hanaguri}}, \bibinfo {author} {\bibfnamefont {Y.}~\bibnamefont {Kohsaka}},
  \bibinfo {author} {\bibfnamefont {M.}~\bibnamefont {Ono}}, \bibinfo {author}
  {\bibfnamefont {M.}~\bibnamefont {Maltseva}}, \bibinfo {author}
  {\bibfnamefont {P.}~\bibnamefont {Coleman}}, \bibinfo {author} {\bibfnamefont
  {I.}~\bibnamefont {Yamada}}, \bibinfo {author} {\bibfnamefont
  {M.}~\bibnamefont {Azuma}}, \bibinfo {author} {\bibfnamefont
  {M.}~\bibnamefont {Takano}}, \bibinfo {author} {\bibfnamefont
  {K.}~\bibnamefont {Ohishi}}, \ and\ \bibinfo {author} {\bibfnamefont
  {H.}~\bibnamefont {Takagi}},\ }\href {\doibase 10.1126/science.1166138}
  {\bibfield  {journal} {\bibinfo  {journal} {Science}\ }\textbf {\bibinfo
  {volume} {323}},\ \bibinfo {pages} {923} (\bibinfo {year} {2009})},\ \Eprint
  {http://arxiv.org/abs/http://www.sciencemag.org/content/323/5916/923.full.pdf}
  {http://www.sciencemag.org/content/323/5916/923.full.pdf} \BibitemShut
  {NoStop}%
\bibitem [{\citenamefont {Hanaguri}\ \emph {et~al.}(2010)\citenamefont
  {Hanaguri}, \citenamefont {Niitaka}, \citenamefont {Kuroki},\ and\
  \citenamefont {Takagi}}]{Hanaguri2010}%
  \BibitemOpen
  \bibfield  {author} {\bibinfo {author} {\bibfnamefont {T.}~\bibnamefont
  {Hanaguri}}, \bibinfo {author} {\bibfnamefont {S.}~\bibnamefont {Niitaka}},
  \bibinfo {author} {\bibfnamefont {K.}~\bibnamefont {Kuroki}}, \ and\ \bibinfo
  {author} {\bibfnamefont {H.}~\bibnamefont {Takagi}},\ }\href {\doibase
  10.1126/science.1187399} {\bibfield  {journal} {\bibinfo  {journal}
  {Science}\ }\textbf {\bibinfo {volume} {328}},\ \bibinfo {pages} {474}
  (\bibinfo {year} {2010})},\ \Eprint
  {http://arxiv.org/abs/http://www.sciencemag.org/content/328/5977/474.full.pdf}
  {http://www.sciencemag.org/content/328/5977/474.full.pdf} \BibitemShut
  {NoStop}%
\bibitem [{\citenamefont {{Wang}}\ \emph {et~al.}(2010)\citenamefont {{Wang}},
  \citenamefont {{Zhai}},\ and\ \citenamefont {{Lee}}}]{Wang2010}%
  \BibitemOpen
  \bibfield  {author} {\bibinfo {author} {\bibfnamefont {F.}~\bibnamefont
  {{Wang}}}, \bibinfo {author} {\bibfnamefont {H.}~\bibnamefont {{Zhai}}}, \
  and\ \bibinfo {author} {\bibfnamefont {D.}~\bibnamefont {{Lee}}},\
  }\href@noop {} {\bibfield  {journal} {\bibinfo  {journal} {\prb}\ }\textbf
  {\bibinfo {volume} {81}},\ \bibinfo {pages} {184512} (\bibinfo {year}
  {2010})}\BibitemShut {NoStop}%
\bibitem [{\citenamefont {Akbari}\ \emph {et~al.}(2010)\citenamefont {Akbari},
  \citenamefont {Knolle}, \citenamefont {Eremin},\ and\ \citenamefont
  {Moessner}}]{Akbari2010}%
  \BibitemOpen
  \bibfield  {author} {\bibinfo {author} {\bibfnamefont {A.}~\bibnamefont
  {Akbari}}, \bibinfo {author} {\bibfnamefont {J.}~\bibnamefont {Knolle}},
  \bibinfo {author} {\bibfnamefont {I.}~\bibnamefont {Eremin}}, \ and\ \bibinfo
  {author} {\bibfnamefont {R.}~\bibnamefont {Moessner}},\ }\href {\doibase
  10.1103/PhysRevB.82.224506} {\bibfield  {journal} {\bibinfo  {journal} {Phys.
  Rev. B}\ }\textbf {\bibinfo {volume} {82}},\ \bibinfo {pages} {224506}
  (\bibinfo {year} {2010})}\BibitemShut {NoStop}%
\bibitem [{\citenamefont {Chi}\ \emph {et~al.}(2014)\citenamefont {Chi},
  \citenamefont {Johnston}, \citenamefont {Levy}, \citenamefont {Grothe},
  \citenamefont {Szedlak}, \citenamefont {Ludbrook}, \citenamefont {Liang},
  \citenamefont {Dosanjh}, \citenamefont {Burke}, \citenamefont {Damascelli},
  \citenamefont {Bonn}, \citenamefont {Hardy},\ and\ \citenamefont
  {Pennec}}]{Chi_etal_2014}%
  \BibitemOpen
  \bibfield  {author} {\bibinfo {author} {\bibfnamefont {S.}~\bibnamefont
  {Chi}}, \bibinfo {author} {\bibfnamefont {S.}~\bibnamefont {Johnston}},
  \bibinfo {author} {\bibfnamefont {G.}~\bibnamefont {Levy}}, \bibinfo {author}
  {\bibfnamefont {S.}~\bibnamefont {Grothe}}, \bibinfo {author} {\bibfnamefont
  {R.}~\bibnamefont {Szedlak}}, \bibinfo {author} {\bibfnamefont
  {B.}~\bibnamefont {Ludbrook}}, \bibinfo {author} {\bibfnamefont
  {R.}~\bibnamefont {Liang}}, \bibinfo {author} {\bibfnamefont
  {P.}~\bibnamefont {Dosanjh}}, \bibinfo {author} {\bibfnamefont {S.~A.}\
  \bibnamefont {Burke}}, \bibinfo {author} {\bibfnamefont {A.}~\bibnamefont
  {Damascelli}}, \bibinfo {author} {\bibfnamefont {D.~A.}\ \bibnamefont
  {Bonn}}, \bibinfo {author} {\bibfnamefont {W.~N.}\ \bibnamefont {Hardy}}, \
  and\ \bibinfo {author} {\bibfnamefont {Y.}~\bibnamefont {Pennec}},\ }\href
  {\doibase 10.1103/PhysRevB.89.104522} {\bibfield  {journal} {\bibinfo
  {journal} {Phys. Rev. B}\ }\textbf {\bibinfo {volume} {89}},\ \bibinfo
  {pages} {104522} (\bibinfo {year} {2014})}\BibitemShut {NoStop}%
\bibitem [{\citenamefont {Kreisel}\ \emph {et~al.}(2015)\citenamefont
  {Kreisel}, \citenamefont {Choubey}, \citenamefont {Berlijn}, \citenamefont
  {Ku}, \citenamefont {Andersen},\ and\ \citenamefont
  {Hirschfeld}}]{Kreisel2015}%
  \BibitemOpen
  \bibfield  {author} {\bibinfo {author} {\bibfnamefont {A.}~\bibnamefont
  {Kreisel}}, \bibinfo {author} {\bibfnamefont {P.}~\bibnamefont {Choubey}},
  \bibinfo {author} {\bibfnamefont {T.}~\bibnamefont {Berlijn}}, \bibinfo
  {author} {\bibfnamefont {W.}~\bibnamefont {Ku}}, \bibinfo {author}
  {\bibfnamefont {B.~M.}\ \bibnamefont {Andersen}}, \ and\ \bibinfo {author}
  {\bibfnamefont {P.~J.}\ \bibnamefont {Hirschfeld}},\ }\href {\doibase
  10.1103/PhysRevLett.114.217002} {\bibfield  {journal} {\bibinfo  {journal}
  {Phys. Rev. Lett.}\ }\textbf {\bibinfo {volume} {114}},\ \bibinfo {pages}
  {217002} (\bibinfo {year} {2015})}\BibitemShut {NoStop}%
\bibitem [{\citenamefont {Capriotti}\ \emph {et~al.}(2003)\citenamefont
  {Capriotti}, \citenamefont {Scalapino},\ and\ \citenamefont
  {Sedgewick}}]{Capriotti03}%
  \BibitemOpen
  \bibfield  {author} {\bibinfo {author} {\bibfnamefont {L.}~\bibnamefont
  {Capriotti}}, \bibinfo {author} {\bibfnamefont {D.~J.}\ \bibnamefont
  {Scalapino}}, \ and\ \bibinfo {author} {\bibfnamefont {R.~D.}\ \bibnamefont
  {Sedgewick}},\ }\href {\doibase 10.1103/PhysRevB.68.014508} {\bibfield
  {journal} {\bibinfo  {journal} {Phys. Rev. B}\ }\textbf {\bibinfo {volume}
  {68}},\ \bibinfo {pages} {014508} (\bibinfo {year} {2003})}\BibitemShut
  {NoStop}%
\bibitem [{\citenamefont {Zhu}\ \emph {et~al.}(2004)\citenamefont {Zhu},
  \citenamefont {Atkinson},\ and\ \citenamefont {Hirschfeld}}]{ZAH04}%
  \BibitemOpen
  \bibfield  {author} {\bibinfo {author} {\bibfnamefont {L.}~\bibnamefont
  {Zhu}}, \bibinfo {author} {\bibfnamefont {W.~A.}\ \bibnamefont {Atkinson}}, \
  and\ \bibinfo {author} {\bibfnamefont {P.~J.}\ \bibnamefont {Hirschfeld}},\
  }\href {\doibase 10.1103/PhysRevB.69.060503} {\bibfield  {journal} {\bibinfo
  {journal} {Phys. Rev. B}\ }\textbf {\bibinfo {volume} {69}},\ \bibinfo
  {pages} {060503} (\bibinfo {year} {2004})}\BibitemShut {NoStop}%
\bibitem [{\citenamefont {Sprunger}\ \emph {et~al.}(1997)\citenamefont
  {Sprunger}, \citenamefont {Petersen}, \citenamefont {Plummer}, \citenamefont
  {Lægsgaard},\ and\ \citenamefont {Besenbacher}}]{Sprunger97}%
  \BibitemOpen
  \bibfield  {author} {\bibinfo {author} {\bibfnamefont {P.~T.}\ \bibnamefont
  {Sprunger}}, \bibinfo {author} {\bibfnamefont {L.}~\bibnamefont {Petersen}},
  \bibinfo {author} {\bibfnamefont {E.~W.}\ \bibnamefont {Plummer}}, \bibinfo
  {author} {\bibfnamefont {E.}~\bibnamefont {Lægsgaard}}, \ and\ \bibinfo
  {author} {\bibfnamefont {F.}~\bibnamefont {Besenbacher}},\ }\href {\doibase
  10.1126/science.275.5307.1764} {\bibfield  {journal} {\bibinfo  {journal}
  {Science}\ }\textbf {\bibinfo {volume} {275}},\ \bibinfo {pages} {1764}
  (\bibinfo {year} {1997})},\ \Eprint
  {http://arxiv.org/abs/http://www.sciencemag.org/content/275/5307/1764.full.pdf}
  {http://www.sciencemag.org/content/275/5307/1764.full.pdf} \BibitemShut
  {NoStop}%
\bibitem [{\citenamefont {Hoffman}\ \emph {et~al.}(2002)\citenamefont
  {Hoffman}, \citenamefont {McElroy}, \citenamefont {Lee}, \citenamefont
  {Lang}, \citenamefont {Eisaki}, \citenamefont {Uchida},\ and\ \citenamefont
  {Davis}}]{Hoffman02}%
  \BibitemOpen
  \bibfield  {author} {\bibinfo {author} {\bibfnamefont {J.~E.}\ \bibnamefont
  {Hoffman}}, \bibinfo {author} {\bibfnamefont {K.}~\bibnamefont {McElroy}},
  \bibinfo {author} {\bibfnamefont {D.~H.}\ \bibnamefont {Lee}}, \bibinfo
  {author} {\bibfnamefont {K.~M.}\ \bibnamefont {Lang}}, \bibinfo {author}
  {\bibfnamefont {H.}~\bibnamefont {Eisaki}}, \bibinfo {author} {\bibfnamefont
  {S.}~\bibnamefont {Uchida}}, \ and\ \bibinfo {author} {\bibfnamefont {J.~C.}\
  \bibnamefont {Davis}},\ }\href {http://www.jstor.org/stable/3832213}
  {\bibfield  {journal} {\bibinfo  {journal} {Science}\ }\bibinfo {series} {New
  Series},\ \textbf {\bibinfo {volume} {297}},\ \bibinfo {pages} {pp. 1148}
  (\bibinfo {year} {2002})}\BibitemShut {NoStop}%
\bibitem [{\citenamefont {Sykora}\ and\ \citenamefont
  {Coleman}(2011)}]{sykora}%
  \BibitemOpen
  \bibfield  {author} {\bibinfo {author} {\bibfnamefont {S.}~\bibnamefont
  {Sykora}}\ and\ \bibinfo {author} {\bibfnamefont {P.}~\bibnamefont
  {Coleman}},\ }\href {\doibase http://dx.doi.org/10.1103/PhysRevB.84.054501}
  {\bibfield  {journal} {\bibinfo  {journal} {Phys. Rev. B}\ }\textbf {\bibinfo
  {volume} {84}},\ \bibinfo {pages} {054501} (\bibinfo {year}
  {2011})}\BibitemShut {NoStop}%
\bibitem [{\citenamefont {H\"anke}\ \emph {et~al.}(2012)\citenamefont
  {H\"anke}, \citenamefont {Sykora}, \citenamefont {Schlegel}, \citenamefont
  {Baumann}, \citenamefont {Harnagea}, \citenamefont {Wurmehl}, \citenamefont
  {Daghofer}, \citenamefont {B\"uchner}, \citenamefont {van~den Brink},\ and\
  \citenamefont {Hess}}]{Buechner11}%
  \BibitemOpen
  \bibfield  {author} {\bibinfo {author} {\bibfnamefont {T.}~\bibnamefont
  {H\"anke}}, \bibinfo {author} {\bibfnamefont {S.}~\bibnamefont {Sykora}},
  \bibinfo {author} {\bibfnamefont {R.}~\bibnamefont {Schlegel}}, \bibinfo
  {author} {\bibfnamefont {D.}~\bibnamefont {Baumann}}, \bibinfo {author}
  {\bibfnamefont {L.}~\bibnamefont {Harnagea}}, \bibinfo {author}
  {\bibfnamefont {S.}~\bibnamefont {Wurmehl}}, \bibinfo {author} {\bibfnamefont
  {M.}~\bibnamefont {Daghofer}}, \bibinfo {author} {\bibfnamefont
  {B.}~\bibnamefont {B\"uchner}}, \bibinfo {author} {\bibfnamefont
  {J.}~\bibnamefont {van~den Brink}}, \ and\ \bibinfo {author} {\bibfnamefont
  {C.}~\bibnamefont {Hess}},\ }\href {\doibase 10.1103/PhysRevLett.108.127001}
  {\bibfield  {journal} {\bibinfo  {journal} {Phys. Rev. Lett.}\ }\textbf
  {\bibinfo {volume} {108}},\ \bibinfo {pages} {127001} (\bibinfo {year}
  {2012})}\BibitemShut {NoStop}%
\bibitem [{\citenamefont {Wang}\ \emph {et~al.}(2009)\citenamefont {Wang},
  \citenamefont {Zhai},\ and\ \citenamefont {Lee}}]{wang}%
  \BibitemOpen
  \bibfield  {author} {\bibinfo {author} {\bibfnamefont {F.}~\bibnamefont
  {Wang}}, \bibinfo {author} {\bibfnamefont {H.}~\bibnamefont {Zhai}}, \ and\
  \bibinfo {author} {\bibfnamefont {D.~H.}\ \bibnamefont {Lee}},\ }\href@noop
  {} {\bibfield  {journal} {\bibinfo  {journal} {EPL}\ }\textbf {\bibinfo
  {volume} {85}},\ \bibinfo {pages} {37005} (\bibinfo {year}
  {2009})}\BibitemShut {NoStop}%
\bibitem [{\citenamefont {Plamadeala}\ \emph {et~al.}(2010)\citenamefont
  {Plamadeala}, \citenamefont {Pereg-Barnea},\ and\ \citenamefont
  {Refael}}]{Plamadela}%
  \BibitemOpen
  \bibfield  {author} {\bibinfo {author} {\bibfnamefont {E.}~\bibnamefont
  {Plamadeala}}, \bibinfo {author} {\bibfnamefont {T.}~\bibnamefont
  {Pereg-Barnea}}, \ and\ \bibinfo {author} {\bibfnamefont {G.}~\bibnamefont
  {Refael}},\ }\href@noop {} {\bibfield  {journal} {\bibinfo  {journal} {Phys.
  Rev. B}\ }\textbf {\bibinfo {volume} {81}},\ \bibinfo {pages} {134513}
  (\bibinfo {year} {2010})}\BibitemShut {NoStop}%
\bibitem [{\citenamefont {Zhang}\ and\ \citenamefont {et~al.}(2010)}]{zhang09}%
  \BibitemOpen
  \bibfield  {author} {\bibinfo {author} {\bibfnamefont {Y.~Y.}\ \bibnamefont
  {Zhang}}\ and\ \bibinfo {author} {\bibnamefont {et~al.}},\ }\href@noop {}
  {\bibfield  {journal} {\bibinfo  {journal} {Phys. Rev. B}\ }\textbf {\bibinfo
  {volume} {80}},\ \bibinfo {pages} {094528} (\bibinfo {year}
  {2010})}\BibitemShut {NoStop}%
\bibitem [{\citenamefont {Wang}\ and\ \citenamefont {Lee}(2003)}]{WangLee03}%
  \BibitemOpen
  \bibfield  {author} {\bibinfo {author} {\bibfnamefont {Q.-H.}\ \bibnamefont
  {Wang}}\ and\ \bibinfo {author} {\bibfnamefont {D.-H.}\ \bibnamefont {Lee}},\
  }\href {\doibase 10.1103/PhysRevB.67.020511} {\bibfield  {journal} {\bibinfo
  {journal} {Phys. Rev. B}\ }\textbf {\bibinfo {volume} {67}},\ \bibinfo
  {pages} {020511} (\bibinfo {year} {2003})}\BibitemShut {NoStop}%
\bibitem [{\citenamefont {Maltseva}\ and\ \citenamefont
  {Coleman}(2009)}]{Maltseva2009}%
  \BibitemOpen
  \bibfield  {author} {\bibinfo {author} {\bibfnamefont {M.}~\bibnamefont
  {Maltseva}}\ and\ \bibinfo {author} {\bibfnamefont {P.}~\bibnamefont
  {Coleman}},\ }\href {\doibase 10.1103/PhysRevB.80.144514} {\bibfield
  {journal} {\bibinfo  {journal} {Phys. Rev. B}\ }\textbf {\bibinfo {volume}
  {80}},\ \bibinfo {pages} {144514} (\bibinfo {year} {2009})}\BibitemShut
  {NoStop}%
\bibitem [{\citenamefont {Efremov}\ \emph {et~al.}(2011)\citenamefont
  {Efremov}, \citenamefont {Korshunov}, \citenamefont {Dolgov}, \citenamefont
  {Golubov},\ and\ \citenamefont {Hirschfeld}}]{Efremov11}%
  \BibitemOpen
  \bibfield  {author} {\bibinfo {author} {\bibfnamefont {D.~V.}\ \bibnamefont
  {Efremov}}, \bibinfo {author} {\bibfnamefont {M.~M.}\ \bibnamefont
  {Korshunov}}, \bibinfo {author} {\bibfnamefont {O.~V.}\ \bibnamefont
  {Dolgov}}, \bibinfo {author} {\bibfnamefont {A.~A.}\ \bibnamefont {Golubov}},
  \ and\ \bibinfo {author} {\bibfnamefont {P.~J.}\ \bibnamefont {Hirschfeld}},\
  }\href {\doibase 10.1103/PhysRevB.84.180512} {\bibfield  {journal} {\bibinfo
  {journal} {Phys. Rev. B}\ }\textbf {\bibinfo {volume} {84}},\ \bibinfo
  {pages} {180512} (\bibinfo {year} {2011})}\BibitemShut {NoStop}%
\bibitem [{Note1()}]{Note1}%
  \BibitemOpen
  \bibinfo {note} {In a $d$-wave superconductor, the large impurity potential
  limit yields $\protect \mathaccentV {hat}05E{t}\propto \tau _{0}$, but in an
  $s$-wave system the nonvanishing integrated anomalous Green's functions
  generate $\tau _{1}$ and $\tau _{2}$ terms as well.}\BibitemShut {Stop}%
\bibitem [{\citenamefont {Hoffman}(2011)}]{Hoffman_rev}%
  \BibitemOpen
  \bibfield  {author} {\bibinfo {author} {\bibfnamefont {J.~E.}\ \bibnamefont
  {Hoffman}},\ }\href {http://stacks.iop.org/0034-4885/74/i=12/a=124513}
  {\bibfield  {journal} {\bibinfo  {journal} {Reports on Progress in Physics}\
  }\textbf {\bibinfo {volume} {74}},\ \bibinfo {pages} {124513} (\bibinfo
  {year} {2011})}\BibitemShut {NoStop}%
\bibitem [{\citenamefont {Beaird}\ \emph {et~al.}(2012)\citenamefont {Beaird},
  \citenamefont {Vekhter},\ and\ \citenamefont {Zhu}}]{Beaird2012}%
  \BibitemOpen
  \bibfield  {author} {\bibinfo {author} {\bibfnamefont {R.}~\bibnamefont
  {Beaird}}, \bibinfo {author} {\bibfnamefont {I.}~\bibnamefont {Vekhter}}, \
  and\ \bibinfo {author} {\bibfnamefont {J.-X.}\ \bibnamefont {Zhu}},\ }\href
  {\doibase 10.1103/PhysRevB.86.140507} {\bibfield  {journal} {\bibinfo
  {journal} {Phys. Rev. B}\ }\textbf {\bibinfo {volume} {86}},\ \bibinfo
  {pages} {140507} (\bibinfo {year} {2012})}\BibitemShut {NoStop}%
\bibitem [{\citenamefont {Derry}\ and\ \citenamefont {Logan}(2015)}]{Derry15}%
  \BibitemOpen
  \bibfield  {author} {\bibinfo {author} {\bibfnamefont {M.~A.}\ \bibnamefont
  {Derry}, \bibfnamefont {P.G.}}\ and\ \bibinfo {author} {\bibfnamefont
  {D.}~\bibnamefont {Logan}},\ }\href@noop {} {\bibfield  {journal} {\bibinfo
  {journal} {arXiv:}\ }\textbf {\bibinfo {volume} {1503.04712}} (\bibinfo
  {year} {2015})}\BibitemShut {NoStop}%
\bibitem [{\citenamefont {Hettler}\ and\ \citenamefont
  {Hirschfeld}(1999)}]{HettlerHirschfeld1999}%
  \BibitemOpen
  \bibfield  {author} {\bibinfo {author} {\bibfnamefont {M.~H.}\ \bibnamefont
  {Hettler}}\ and\ \bibinfo {author} {\bibfnamefont {P.~J.}\ \bibnamefont
  {Hirschfeld}},\ }\href {\doibase 10.1103/PhysRevB.59.9606} {\bibfield
  {journal} {\bibinfo  {journal} {Phys. Rev. B}\ }\textbf {\bibinfo {volume}
  {59}},\ \bibinfo {pages} {9606} (\bibinfo {year} {1999})}\BibitemShut
  {NoStop}%
\bibitem [{\citenamefont {Shnirman}\ \emph {et~al.}(1999)\citenamefont
  {Shnirman}, \citenamefont {Adagideli}, \citenamefont {Goldbart},\ and\
  \citenamefont {Yazdani}}]{Shnirman1999}%
  \BibitemOpen
  \bibfield  {author} {\bibinfo {author} {\bibfnamefont {A.}~\bibnamefont
  {Shnirman}}, \bibinfo {author} {\bibfnamefont {i.~d.~I.}\ \bibnamefont
  {Adagideli}}, \bibinfo {author} {\bibfnamefont {P.~M.}\ \bibnamefont
  {Goldbart}}, \ and\ \bibinfo {author} {\bibfnamefont {A.}~\bibnamefont
  {Yazdani}},\ }\href {\doibase 10.1103/PhysRevB.60.7517} {\bibfield  {journal}
  {\bibinfo  {journal} {Phys. Rev. B}\ }\textbf {\bibinfo {volume} {60}},\
  \bibinfo {pages} {7517} (\bibinfo {year} {1999})}\BibitemShut {NoStop}%
\bibitem [{\citenamefont {Khlyustikov}\ and\ \citenamefont
  {Buzdin}(1987)}]{Buzdin87}%
  \BibitemOpen
  \bibfield  {author} {\bibinfo {author} {\bibfnamefont {I.~N.}\ \bibnamefont
  {Khlyustikov}}\ and\ \bibinfo {author} {\bibfnamefont {A.~I.}\ \bibnamefont
  {Buzdin}},\ }\href@noop {} {\bibfield  {journal} {\bibinfo  {journal} {Adv.
  Phys.}\ }\textbf {\bibinfo {volume} {36}},\ \bibinfo {pages} {271} (\bibinfo
  {year} {1987})}\BibitemShut {NoStop}%
\bibitem [{\citenamefont {Nunner}\ \emph {et~al.}(2005)\citenamefont {Nunner},
  \citenamefont {Andersen}, \citenamefont {Melikyan},\ and\ \citenamefont
  {Hirschfeld}}]{Nunner2005}%
  \BibitemOpen
  \bibfield  {author} {\bibinfo {author} {\bibfnamefont {T.~S.}\ \bibnamefont
  {Nunner}}, \bibinfo {author} {\bibfnamefont {B.~M.}\ \bibnamefont
  {Andersen}}, \bibinfo {author} {\bibfnamefont {A.}~\bibnamefont {Melikyan}},
  \ and\ \bibinfo {author} {\bibfnamefont {P.~J.}\ \bibnamefont {Hirschfeld}},\
  }\href {\doibase 10.1103/PhysRevLett.95.177003} {\bibfield  {journal}
  {\bibinfo  {journal} {Phys. Rev. Lett.}\ }\textbf {\bibinfo {volume} {95}},\
  \bibinfo {pages} {177003} (\bibinfo {year} {2005})}\BibitemShut {NoStop}%
\bibitem [{\citenamefont {Ma\ifmmode~\acute{s}\else \'{s}\fi{}ka}\ \emph
  {et~al.}(2007)\citenamefont {Ma\ifmmode~\acute{s}\else \'{s}\fi{}ka},
  \citenamefont {\ifmmode \acute{S}\else \'{S}\fi{}led\ifmmode~\acute{z}\else
  \'{z}\fi{}}, \citenamefont {Czajka},\ and\ \citenamefont
  {Mierzejewski}}]{Maska2007}%
  \BibitemOpen
  \bibfield  {author} {\bibinfo {author} {\bibfnamefont {M.~M.}\ \bibnamefont
  {Ma\ifmmode~\acute{s}\else \'{s}\fi{}ka}}, \bibinfo {author} {\bibfnamefont
  {i.~d.~Z.}\ \bibnamefont {\ifmmode \acute{S}\else
  \'{S}\fi{}led\ifmmode~\acute{z}\else \'{z}\fi{}}}, \bibinfo {author}
  {\bibfnamefont {K.}~\bibnamefont {Czajka}}, \ and\ \bibinfo {author}
  {\bibfnamefont {M.}~\bibnamefont {Mierzejewski}},\ }\href {\doibase
  10.1103/PhysRevLett.99.147006} {\bibfield  {journal} {\bibinfo  {journal}
  {Phys. Rev. Lett.}\ }\textbf {\bibinfo {volume} {99}},\ \bibinfo {pages}
  {147006} (\bibinfo {year} {2007})}\BibitemShut {NoStop}%
\bibitem [{\citenamefont {Foyevtsova}\ \emph {et~al.}(2009)\citenamefont
  {Foyevtsova}, \citenamefont {Valent\'{i}},\ and\ \citenamefont
  {Hirschfeld}}]{Foyevtsova09}%
  \BibitemOpen
  \bibfield  {author} {\bibinfo {author} {\bibfnamefont {K.}~\bibnamefont
  {Foyevtsova}}, \bibinfo {author} {\bibfnamefont {R.}~\bibnamefont
  {Valent\'{i}}}, \ and\ \bibinfo {author} {\bibfnamefont {P.~J.}\ \bibnamefont
  {Hirschfeld}},\ }\href {\doibase 10.1103/PhysRevB.79.144424} {\bibfield
  {journal} {\bibinfo  {journal} {Phys. Rev. B}\ }\textbf {\bibinfo {volume}
  {79}},\ \bibinfo {pages} {144424} (\bibinfo {year} {2009})}\BibitemShut
  {NoStop}%
\bibitem [{\citenamefont {Foyevtsova}\ \emph {et~al.}(2010)\citenamefont
  {Foyevtsova}, \citenamefont {Kandpal}, \citenamefont {Jeschke}, \citenamefont
  {Graser}, \citenamefont {Cheng}, \citenamefont {Valent\'{i}},\ and\
  \citenamefont {Hirschfeld}}]{Foyevtsova10}%
  \BibitemOpen
  \bibfield  {author} {\bibinfo {author} {\bibfnamefont {K.}~\bibnamefont
  {Foyevtsova}}, \bibinfo {author} {\bibfnamefont {H.~C.}\ \bibnamefont
  {Kandpal}}, \bibinfo {author} {\bibfnamefont {H.~O.}\ \bibnamefont
  {Jeschke}}, \bibinfo {author} {\bibfnamefont {S.}~\bibnamefont {Graser}},
  \bibinfo {author} {\bibfnamefont {H.-P.}\ \bibnamefont {Cheng}}, \bibinfo
  {author} {\bibfnamefont {R.}~\bibnamefont {Valent\'{i}}}, \ and\ \bibinfo
  {author} {\bibfnamefont {P.~J.}\ \bibnamefont {Hirschfeld}},\ }\href
  {\doibase 10.1103/PhysRevB.82.054514} {\bibfield  {journal} {\bibinfo
  {journal} {Phys. Rev. B}\ }\textbf {\bibinfo {volume} {82}},\ \bibinfo
  {pages} {054514} (\bibinfo {year} {2010})}\BibitemShut {NoStop}%
\bibitem [{\citenamefont {Yang}\ \emph {et~al.}(2011)\citenamefont {Yang},
  \citenamefont {Lu}, \citenamefont {Xiang},\ and\ \citenamefont
  {Lee}}]{FWang_dwave_2011}%
  \BibitemOpen
  \bibfield  {author} {\bibinfo {author} {\bibfnamefont {M.}~\bibnamefont
  {Yang}, \bibfnamefont {F.and~Gao}}, \bibinfo {author} {\bibfnamefont {Z.-Y.}\
  \bibnamefont {Lu}}, \bibinfo {author} {\bibfnamefont {T.}~\bibnamefont
  {Xiang}}, \ and\ \bibinfo {author} {\bibfnamefont {D.-H.}\ \bibnamefont
  {Lee}},\ }\href@noop {} {\bibfield  {journal} {\bibinfo  {journal} {Europhys.
  Lett.}\ }\textbf {\bibinfo {volume} {93}},\ \bibinfo {pages} {57003}
  (\bibinfo {year} {2011})}\BibitemShut {NoStop}%
\bibitem [{\citenamefont {Maier}\ \emph {et~al.}(2011)\citenamefont {Maier},
  \citenamefont {Graser}, \citenamefont {Hirschfeld},\ and\ \citenamefont
  {Scalapino}}]{Graser_dwave_2011}%
  \BibitemOpen
  \bibfield  {author} {\bibinfo {author} {\bibfnamefont {T.}~\bibnamefont
  {Maier}}, \bibinfo {author} {\bibfnamefont {S.}~\bibnamefont {Graser}},
  \bibinfo {author} {\bibfnamefont {P.}~\bibnamefont {Hirschfeld}}, \ and\
  \bibinfo {author} {\bibfnamefont {D.}~\bibnamefont {Scalapino}},\ }\href@noop
  {} {\bibfield  {journal} {\bibinfo  {journal} {Phys. Rev. B}\ }\textbf
  {\bibinfo {volume} {83}},\ \bibinfo {pages} {100515} (\bibinfo {year}
  {2011})}\BibitemShut {NoStop}%
\bibitem [{\citenamefont {Yin}\ \emph {et~al.}(2014)\citenamefont {Yin},
  \citenamefont {Haule},\ and\ \citenamefont {Kotliar}}]{Yin2014}%
  \BibitemOpen
  \bibfield  {author} {\bibinfo {author} {\bibfnamefont {Z.~P.}\ \bibnamefont
  {Yin}}, \bibinfo {author} {\bibfnamefont {K.}~\bibnamefont {Haule}}, \ and\
  \bibinfo {author} {\bibfnamefont {G.}~\bibnamefont {Kotliar}},\ }\href@noop
  {} {\bibfield  {journal} {\bibinfo  {journal} {Nature Phys.}\ }\textbf
  {\bibinfo {volume} {10}},\ \bibinfo {pages} {845} (\bibinfo {year}
  {2014})}\BibitemShut {NoStop}%
\bibitem [{\citenamefont {Ahn}\ \emph {et~al.}(2014)\citenamefont {Ahn},
  \citenamefont {Eremin}, \citenamefont {Knolle}, \citenamefont {Zabolotnyy},
  \citenamefont {Borisenko}, \citenamefont {B\"uchner},\ and\ \citenamefont
  {Chubukov}}]{Ahn2014}%
  \BibitemOpen
  \bibfield  {author} {\bibinfo {author} {\bibfnamefont {F.}~\bibnamefont
  {Ahn}}, \bibinfo {author} {\bibfnamefont {I.}~\bibnamefont {Eremin}},
  \bibinfo {author} {\bibfnamefont {J.}~\bibnamefont {Knolle}}, \bibinfo
  {author} {\bibfnamefont {V.}~\bibnamefont {Zabolotnyy}}, \bibinfo {author}
  {\bibfnamefont {S.}~\bibnamefont {Borisenko}}, \bibinfo {author}
  {\bibfnamefont {B.}~\bibnamefont {B\"uchner}}, \ and\ \bibinfo {author}
  {\bibfnamefont {A.}~\bibnamefont {Chubukov}},\ }\href@noop {} {\bibfield
  {journal} {\bibinfo  {journal} {Phys. Rev. B}\ }\textbf {\bibinfo {volume}
  {89}},\ \bibinfo {pages} {144513} (\bibinfo {year} {2014})}\BibitemShut
  {NoStop}%
\bibitem [{\citenamefont {Tinkham}(2014)}]{Tinkham}%
  \BibitemOpen
  \bibfield  {author} {\bibinfo {author} {\bibfnamefont {M.}~\bibnamefont
  {Tinkham}},\ }\href@noop {} {\bibfield  {journal} {\bibinfo  {journal}
  {Introduction to Superconductivity: Second Edition}\ ,\ \bibinfo {pages}
  {{\it Dover Books in Physics}}} (\bibinfo {year} {2014})}\BibitemShut
  {NoStop}%
\end{thebibliography}%

\begin{widetext}

\section{Appendix}

\subsection{Remarks on the Bragg peaks and the 1-Fe vs. 2-Fe Brillouin zone}

The electronic structure of the iron-based superconductors can be described exactly in the Brillouin
zone corresponding to a 2-Fe unit cell, or approximately in the twice larger
one, corresponding to a 1-Fe cell\cite{HKM_ROPP11}. In the former case, the
intraband scattering create a QPI spot near the zone center
($\mathbf{q\approx0),}$ and an intraband one near zone corners
($\mathbf{q\approx}\pi/a,\pi/a\mathbf{),}$ where $a$ is the lattice parameter
of the 2-Fe cell, $a=d\sqrt{2},$ and $d$ is the Fe-Fe bond length. In the
experiment, the QPI signal is invariably mixed with the Bragg peaks, resulting
from electrons scattering off the regular crystal lattice. It is important to
understand that, just as in X-ray scattering, while the density of states
$\rho(\mathbf{q)}$ is peaked at each reciprocal lattice \textbf{G}, the
intensity of each $\mathbf{q+G}$ component depends on $\mathbf{G,}$ and
generally decays with $|G|,$ and the same is true for the QPI spectra. In
other word, when we say that the intraband spot is located around
$\mathbf{q=0,}$ it is implied that there are also spots near all
$\mathbf{q=G,}$ albeit with a reduced intensity.

Now, let us use the 1-Fe cell, and a twice larger Brillouin zone. In this
setting, there are spots near $\mathbf{q}_{1}\mathbf{\approx0,}$ near
$\mathbf{q}_{2}\mathbf{\approx}\pi/d,0\mathbf{,}$ and two overlapping spots
near $\mathbf{q}_{3,4}\mathbf{\approx}\pi/d,\pi/d$. Three of them, the first one and the last two, are located at
2-Fe reciprocal lattice vectors. Of course, this means that as long as we
include scattering off the pnictogen or chalcogen sublattice, each of these
peaks will generate shadow peaks at all other reciprocal lattice vectors.
Experimentally, however, they will be clearly distinguishable: the
\textbf{q}$_{1}$ peak will be most intensive at $\mathbf{G=0,}$ while
$\mathbf{q}_{3,4}$ will actually be stronger at $\mathbf{G}=0\mathbf{,}\pm
\pi/a$ and $\pm\pi/a,0$ (in the reduced zone). The spots near $\mathbf{q}_{2}$
will not be affected by the downfolding procedure, as for this particular
vector the sublattice scattering will not generate any shadows.

This may seem to be of academic importance, but it may have considerable
practical ramifications. Indeed, while for Fe-based superconductors, as mentioned, there is no
problem separating the hole-hole and electron-electron scattering from the
hole-electron one, the scattering between two inequivalent electron pockets in
the 2-Fe zone is seemingly indistinguishable from the intraband scattering,
and, as the latter, overlaps with the Bragg peaks. But, as discussed above,
the intraband scattering will be stronger at that half of the reciprocal
lattice vectors that coincide with those reciprocal lattice vectors of the 1-Fe
cell, while the interband electron-electron scattering will be stronger at the
other half of the \textbf{G}-vectors. This, of course, makes the signatures of
the order parameter signs discussed in the main text weaker, but does not
destroy them. Moreover, several papers recently suggested a possibility of the
order parameter sign change between different hole bands\cite{Yin2014,Ahn2014} in
particular, between the two bands formed by the $xz$ and $yz$ orbitals and the
one with predominantly $xy$ character. In the 2-Fe zone they both occur at the
zone center and thus seem indistinguishable. However, a closer look reveals
that the first two bands generate a spot that is located at the zone center in
both 1-Fe and 2-Fe zones, while the scattering between $xz/yz$ and $xy$
pockets, in the 1-Fe zone occurs near the zone corner. While both processes
will create \textquotedblleft shadows\textquotedblright\ at all 2-Fe
reciprocal lattice vectors, the former will be stronger at one half of the
vectors, and the other at the other half, again allowing to use QPI to assess
the above hypothesis.

Finally, the very fact that some of the nontrivial QPI spots are overlapping
with Bragg peaks is not a disadvantage, but just the opposite. As pointed out
by Hanaguri \textit{et al.}\cite{Hanaguri2010}, the Bragg peaks are much sharper
then the QPI maxima, and have distinctly different profiles, which allows them
to be separated. Note that the mechanism that is supposed to generate a QPI
dependence on the external magnetic field should not operate on the Bragg
peaks, whether in the normal or superconducting state. Thus, if not only the
QPI, but also the Bragg peaks demonstrate a strong field dependence (as was in
fact the case in Ref. \cite{Hanaguri2010}), this strongly suggests that the field
dependence of the QPI spectra is also not directly related to scattering off
Abrikosov vortices. Given the caveats described in the main text, this
capability appears rather useful.

\subsection{ Coherence factors}

We remarked in the introduction that there is an expectation shared by many in
the STM community that the temperature and bias dependence of QPI signals will
follow simple BCS coherence factors. This is based in part on Refs.
\onlinecite{Hanaguri2009},\onlinecite{Maltseva2009}, which expressed the
change in the Fourier transformed density of states as an integral over such
factors, written explicitly, for instance, for the antisymmetrized response
and Born scattering, as $(u_{\mathbf{k}}u_{\mathbf{k}^{\prime}}-v_{\mathbf{k}%
}v_{\mathbf{k}^{\prime}})^{2},$ and similar expressions for other types of
scattering. However, this assertion is not correct and the prefactors in the
expressions for the QPI intensity cannot be cast into such form. Let us
illustrate that now for the Born scattering. In this case,
\begin{align}
\delta\rho(\mathbf{q},\omega)  &  =\frac{1}{2}Tr\operatorname{Im}%
\int_{\mathbf{k}}\tau_{3}G_\mathbf{k}(\omega)\tau_{3}G_{\mathbf{k}^{\prime}}%
(\omega)\delta(\mathbf{k-k}^{\prime}\mathbf{-q})\\
&  =\operatorname{Im}\int_{\mathbf{k}}\frac{\omega^{2}+\xi_{\mathbf{k}}%
\xi_{\mathbf{k+q}}-\Delta_{\mathbf{k}}\Delta_{\mathbf{k}^{\prime}}}%
{(\omega^{2}-E_{\mathbf{k}}^{2})(\omega^{2}-E_{\mathbf{k}^{\prime}}^{2}%
)}\delta(\mathbf{k-k}^{\prime}\mathbf{-q}).
\end{align}
where $\xi$ are the one-electron energies. Concentrating on the fraction under
the integral, we see that it is
\begin{align}
&  \operatorname{Im}\frac{\omega^{2}+\xi_\mathbf{k}\xi_{\mathbf{k}}-\Delta
_{\mathbf{k}}\Delta_{\mathbf{k}}}{(\omega^{2}-E_{\mathbf{k}}^{2})(\omega
^{2}-E_{\mathbf{k}^{\prime}}^{2})}=\operatorname{Im}(\omega^{2}+\xi
_{\mathbf{k}}\xi_{\mathbf{k}^{\prime}}-\Delta_\mathbf{k}\Delta_{\mathbf{k}^{\prime}%
}) \nonumber\\
&  \times\Big[\operatorname{Im}\frac{1}{4E_{\mathbf{k}}}\left(  \frac
{1}{\omega-E_{\mathbf{k}}}-\frac{1}{\omega+E_{\mathbf{k}^{\prime}}}\right)
\operatorname{Re}\frac{1}{(\omega^{2}-E_{\mathbf{k}^{\prime}}^{2})} \nonumber\\
&  +\operatorname{Im}\frac{1}{4E_{\mathbf{k}^{\prime}}}\left(  \frac{1}%
{\omega-E_{\mathbf{k}^{\prime}}}-\frac{1}{\omega+E_{\mathbf{k}^{\prime}}%
}\right)  \operatorname{Re}\frac{1}{(\omega^{2}-E_{\mathbf{k}}^{2})}\Big] \nonumber\\
&  =\Big[\frac{\delta(\omega-E_{\mathbf{k}})(E_{\mathbf{k}}^{2}+\xi_{\mathbf{k}%
}\xi_{\mathbf{k}^{\prime}}-\Delta_{\mathbf{k}}\Delta_{\mathbf{k}^{\prime}}%
)}{4E_{\mathbf{k}}(E_{\mathbf{k}}^{2}{}-E_{\mathbf{k}^{\prime}}^{2})} \nonumber\\
&  +\frac{\delta(\omega-E_{\mathbf{k}^{\prime}})(E_{\mathbf{k}^{\prime}}%
^{2}+\xi_{\mathbf{k}}\xi_{\mathbf{k}^{\prime}}-\Delta_{\mathbf{k}}%
\Delta_{\mathbf{k}^{\prime}})}{4E_{\mathbf{k}^{\prime}}(E_{\mathbf{k}^{\prime
}}^{2}-E_{\mathbf{k}}^{2})}\Big] \nonumber\\
&  =\frac{1}{(E_{\mathbf{k}}^{2}{}-E_{\mathbf{k}^{\prime}}^{2})}[\frac
{\delta(\omega-E_{\mathbf{k}})(E_{\mathbf{k}}^{2}+\xi_{\mathbf{k}}%
\xi_{\mathbf{k}^{\prime}}-\Delta_{\mathbf{k}}\Delta_{\mathbf{k}^{\prime}}%
)}{4E_{\mathbf{k}}} \nonumber\\
&  -\frac{\delta(\omega-E_{\mathbf{k}^{\prime}})(E_{\mathbf{k}^{\prime}}%
^{2}+\xi_{\mathbf{k}}\xi_{\mathbf{k}^{\prime}}-\Delta_{\mathbf{k}}%
\Delta_{\mathbf{k}^{\prime}})}{4E_{\mathbf{k}^{\prime}}}\Big]
\end{align}
Note that arguments of the $\delta-$functions are different therefore in order
to combine the two terms we need to rename the $\mathbf{k}$ variables in the
second term, after which, assuming inversion symmetry in $\Delta$ and $\xi$,
we get:
\begin{align}
\delta\rho(\mathbf{q},\omega)  &  =\int_{\mathbf{k}}\Big[\frac{\delta
(\omega-E_{\mathbf{k}})(E_{\mathbf{k}}^{2}+\xi_{\mathbf{k}}\xi_{\mathbf{k}%
^{\prime}}-\Delta_{\mathbf{k}}\Delta_{\mathbf{k}^{\prime}})}{8E_{\mathbf{k}%
}(E_{\mathbf{k}}{}^{2}-E_{\mathbf{k}^{\prime}}^{2})} \nonumber\\
&  +\frac{\delta(\omega-E_{\mathbf{k}^{\prime}})(E_{\mathbf{k}^{\prime}}%
^{2}+\xi_{\mathbf{k}}\xi_{\mathbf{k}^{\prime}}-\Delta_{\mathbf{k}}%
\Delta_{\mathbf{k}^{\prime}})}{4E_{\mathbf{k}^{\prime}}(E_{\mathbf{k}^{\prime
}}^{2}-E_{\mathbf{k}}^{2})}\Big]\delta(\mathbf{k-k}^{\prime}\mathbf{-q}) \nonumber\\
&  =\int_{\mathbf{k}}\frac{(E_{\mathbf{k}}^{2}{}+\xi_{\mathbf{k}}%
\xi_{\mathbf{k}^{\prime}}-\Delta_{\mathbf{k}}\Delta_{\mathbf{k}^{\prime}%
})\delta(\omega-E_{\mathbf{k}})}{(E_{\mathbf{k}}^{2}-E_{\mathbf{k}^{\prime}%
}^{2})\omega}\frac{\delta(\mathbf{k-k}^{\prime}\mathbf{-q})}{2}.
\end{align}
The first factor is what was taken to be a coherence factor in Refs.
\onlinecite{Hanaguri2009} and \onlinecite{Maltseva2009}. However, the true coherence
factor in question is different, namely $(E_{\mathbf{k}}^{2}+\xi_{\mathbf{k}%
}\xi_{\mathbf{k}^{\prime}}-\Delta_{\mathbf{k}}\Delta_{\mathbf{k}^{\prime}%
})/(E_{\mathbf{k}}{}E_{\mathbf{k}^{\prime}}),$ that instead of the expression
above one would have a very different formula, namely%
\begin{align}
&  \int_{\mathbf{k}}\frac{(\omega^{2}+\xi_{\mathbf{k}}\xi_{\mathbf{k}^{\prime
}}-\Delta_{\mathbf{k}}\Delta_{\mathbf{k}^{\prime}})\delta(\omega
-E_{\mathbf{k}})\delta(\omega-E_{\mathbf{k}^{\prime}})}{\omega^{2}}%
\frac{\delta(\mathbf{k-k}^{\prime}\mathbf{-q})}{2}\nonumber\\
&  \neq\int_{\mathbf{k}}\frac{(\omega^{2}+\xi_{\mathbf{k}}\xi_{\mathbf{k}%
^{\prime}}-\Delta_{\mathbf{k}}\Delta_{\mathbf{k}^{\prime}})\delta
(\omega-E_{\mathbf{k}})}{\omega({}\omega^{2}-E_{\mathbf{k}^{\prime}}^{2}%
)}\frac{\delta(\mathbf{k-k}^{\prime}\mathbf{-q})}{2}.
\end{align}

One may think that this difference will disappear after an integration over
\textbf{q,} and the integrated expression will be similar to classic BCS
predictions of thermally averaged transition probabilities, \textit{e.g.} in
NMR spin relaxation and/or ultrasonic attenuation\cite{Tinkham}. For example,
consider the spin-lattice relaxation rate
\begin{align}
{\frac{1}{T_{1}T}}  &  \propto\lim_{\omega_{0}\rightarrow0}\sum_{\mathbf{q,\mu
,\nu}}{\frac{\mathrm{Im}~\chi^{\mu\nu}(\mathbf{q},\omega_{0})}{\omega_{0}}%
}\label{eq:NMR}\\
&  \simeq\lim_{\omega_{0}\rightarrow0}{\frac{1}{\omega_{0}}}\mathrm{Im}%
\sum_{\mathbf{k},\mathbf{k}^{\prime},\mu,\nu}\mathrm{Tr}\left(  {\hat{\alpha}%
}{\hat{G}}^{0}(\mathbf{k},\omega_{n}){\hat{\alpha}}{\hat{G}}^{0}%
(\mathbf{k}^{\prime},\omega_{n}+\omega_{0})\right)  _{i\omega_{0}%
\rightarrow\omega_{0}+i 0^+},\nonumber
\end{align}
where $\chi$ is the spin susceptibility matrix in band space, $\hat{\alpha}$
is the electronic spin operator in Nambu space, $\omega_{0}$ is the Larmor
frequency, and we have neglected any momentum dependence of hyperfine matrix
elements in order to arrive at a simple expression. In fact, Eq. (\ref{eq:NMR})
bears a certain resemblance to the structure of Eq. (\ref{drhoFourier}), in
particular because each Green's function is local and therefore integrated
over momentum independently. Performing these sums and the analytical
continuation, one obtains the usual result, with slight generalization to
multiband systems:%

\begin{align}
{\frac{1}{T_{1}T}}  &  \propto\sum_{\mu,\nu}\int_{0}^{\infty}d\omega\left(
-{\frac{\partial f}{\partial\omega}}\right)  \Big[\mathrm{Im}\left(
{\frac{\omega}{\sqrt{\Delta_{\mu}^{2}-\omega^{2}}}}\right)  \mathrm{Im}\left(
{\frac{\omega}{\sqrt{\Delta_{\nu}^{2}-\omega^{2}}}}\right) \label{eq:T1T}\\
&  +\mathrm{Im}\left(  {\frac{\Delta_{\mu}}{\sqrt{\Delta_{\mu}^{2}-\omega^{2}%
}}}\right)  \mathrm{Im}\left(  {\frac{\Delta_{\nu}}{\sqrt{\Delta_{\nu}%
^{2}-\omega^{2}}}}\right)  \Big] \nonumber\\
&  =\mathrm{intraband~terms}+2\int_{0}^{\infty}d\omega\left(  -{\frac{\partial
f}{\partial\omega}}\right)  \Big[\mathrm{Im}\left(  {\frac{\omega}%
{\sqrt{\Delta_{h}^2-\omega^{2}}}}\right)  \mathrm{Im}\left(  {\frac{\omega
}{\sqrt{\Delta_{e}^2-\omega^{2}}}}\right) \nonumber\\
&  +\mathrm{Im}\left(  {\frac{\Delta_{h}}{\sqrt{\Delta_{h}^2-\omega^{2}}}%
}\right)  \mathrm{Im}\left(  {\frac{\Delta_{e}}{\sqrt{\Delta_{e}^2-\omega^{2}}}%
}\right)  \Big]\nonumber \\
&  =\mathrm{intraband~terms}+2\int_{\max(\Delta_{h},\Delta_{e})}^{\infty
}d\omega\left(  -{\frac{\partial f}{\partial\omega}}\right)  {\frac{\omega
^{2}+\Delta_{h}\Delta_{e}}{\sqrt{\omega^{2}-\Delta_{h}^2}\sqrt{\omega^{2}%
-\Delta_{e}^2}}} \nonumber.
\end{align}

Now one can see that an $s_{++}$ superconductor will have an interband
contribution to NMR relaxation that obeys exactly a BCS type II coherence
factor $T$-dependence (in the limit $\Delta_{h}\rightarrow\Delta_{e}$), while
for $s_{\pm}$ the corresponding result is type I.

Compare now to our expression for the interband contribution to the Fourier
transform density of states in the case of a weak potential scatterer, Eq.
(\ref{eq:rho_inter}), together with the thermal average,
\begin{align}
\langle\delta\rho_{inter}(\omega)\rangle &  \propto\int d\omega\left(
-{\frac{\partial f}{\partial\omega}}\right)  \mathrm{Im}{\frac{\omega
^{2}-\Delta_{h}\Delta_{e}}{\sqrt{\omega^{2}-\Delta_{h}^2}\sqrt{\omega^{2}%
-\Delta_{e}^2}}}\nonumber\label{eq:drho_av}\\
&  =\int_{\Delta_{e}}^{\Delta_{h}}d\omega\left(  -{\frac{\partial f}%
{\partial\omega}}\right)  {\frac{\omega^{2}-\Delta_{h}\Delta_{e}}{\sqrt
{\Delta_{h}^2-\omega^{2}}\sqrt{\omega^{2}-\Delta_{e}^2}}}.
\end{align}
We see that while the functional form of the fraction under the integral is
indeed the same, the cost of the $\mathbf{q-}$integration is that the
frequency integral is now taken over a totally different, in fact not
overlapping, limits,
which results in a completely different $T$-dependence, as shown above.

\vskip .2cm

\subsection{Single impurity $t$-matrix for 2-band system}

The $t$-matrix for a single impurity is given by
\begin{equation}
\underline{{t}}=\left[  1-\underline{{U}}\tau_{3}\sum_{\mathbf{k}%
}{\underline{{G}}(\mathbf{k},\omega)}\right]  ^{-1}{\tau_{3}}\underline{{{U}}}%
\end{equation}
where the integrated matrix Green's function given in Eq.(\ref{integrated_G})
can be further re-written as $\left(  \sum_{\mathbf{k}}\underbar
G(\mathbf{k},\omega)\right)  _{\nu}=g_{\omega,\nu}\tau_{0}+g_{\Delta,\nu
}\tau_{1}$ for each of the band. The scattering matrix \underline{$U$}
can be then separated into intraband $(\underline{U})_{\nu\nu}=U_{intra}$ and interband $(\underline{U})_{\mu\nu}=U_{inter}$ term.

Inverting the $t$ matrix one finds the common denominator for both intra and
interband terms of the $t$ matrix
\begin{eqnarray}
D & = & -\left(  g_{\omega,h}^{2}-g_{\Delta,h}^{2}\right)  U_{intra}^{2}-\left(
g_{\omega,e}^{2}-g_{\Delta,e}^{2}\right)  U_{intra}^{2}-2\left(  g_{\omega
,h}g_{\omega,e}-g_{\Delta,h}g_{\Delta,e}\right)  U_{inter}^{2} \nonumber \\
&& + \left(
g_{\omega,h}^{2}-g_{\Delta,h}^{2}\right)  \left(  g_{\omega,2}^{2}%
-g_{\Delta,2}^{2}\right)  \left(  U_{inter}^{2}-U_{intra}^{2}\right)  {}%
^{2}+1.
\end{eqnarray}

Now for the intraband scattering within the band 1 one obtains in Nambu
components
\begin{equation}
{\hat{t}}_{h}=t_{h}^{0}\tau_{0}+t_{h}^{1}\tau_{1}+t_{h}^{3}\tau_{3}%
\end{equation}
with \begin{equation}
\begin{split}
t_{h}^{0} &  =\left[  g_{\omega,h}U_{intra}^{2}+g_{\omega,e}U_{inter}%
^{2}-g_{\omega,h}\left(  g_{\omega,e}^{2}-g_{\Delta,e}^{2}\right)  \left(
U_{inter}^{2}-U_{intra}^{2}\right)  {}^{2}\right]  /D,\\
t_{h}^{1} &  =\left[  -g_{\Delta,h}U_{intra}^{2}-g_{\Delta,e}U_{inter}%
^{2}+g_{\Delta,h}\left(  g_{\omega,e}^{2}-g_{\Delta,e}^{2}\right)  \left(
U_{inter}^{2}-U_{intra}^{2}\right)  {}^{2}\right]  /D,\\
t_{h}^{3} &  =\left[  U_{intra}\left(  1-\left(  g_{\omega,e}^{2}-g_{\Delta
,e}^{2}\right)  \left(  U_{intra}^{2}-U_{inter}^{2}\right)  \right)  \right]
/D.
\end{split}
\end{equation}
 The expressions for $\hat{t}_{e}$ are obtained by exchanging the band
indices $h$ and $e$. Similarly, the interband $t$-matrix can be written as
\begin{equation}
{\hat{t}}_{eh}=t_{eh}^{0}\tau_{0}+t_{eh}^{1}\tau_{1}+t_{eh}^{2}\tau_{2}%
+t_{eh}^{3}\tau_{3}%
\end{equation}
with
\begin{equation}%
\begin{split}
t_{eh}^{0} &  =\left[  \left(  g_{\omega,e}+g_{\omega,h}\right)
U_{intra}U_{inter}\right]  /D\\
t_{eh}^{1} &  =\left[  -\left(  g_{\Delta,e}+g_{\Delta,h}\right)
U_{intra}U_{inter}\right]  /D\\
t_{eh}^{2} &  =\left[  -i\left(  g_{\Delta,e}g_{\omega,h}-g_{\omega
,e}g_{\Delta,h}\right)  U_{inter}\left(  U_{inter}^{2}-U_{intra}^{2}\right)
\right]  /D\\
t_{eh}^{3} &  =\left[  U_{inter}\left(  g_{\omega,e}g_{\omega,h}-g_{\Delta
,e}g_{\Delta,h}\right)  \left(  U_{intra}^{2}-U_{inter}^{2}\right)  +1\right]
/D,
\end{split}
\end{equation}
and the corresponding expression for $t_{he}.$ Note that the interband
scattering terms generate a $\tau_{2}$ contribution for the interband
$t$-matrix.

\end{widetext}

\end{document}